\documentclass[11pt]{article}
\usepackage[utf8]{inputenc}
\usepackage{fontawesome}

\usepackage{stmaryrd}
\usepackage{varioref}
\usepackage{amsmath,amsfonts,amsthm,mathrsfs}
\usepackage{mathabx}
\usepackage[normalem]{ulem}
\usepackage{multirow,color,graphics}
\usepackage{amsmath}
\usepackage{amsfonts}
\usepackage{amssymb}
\usepackage{jheppub}
\usepackage{fancyvrb}
\usepackage{verbatim}
\usepackage{wrapfig}
\usepackage{appendix}
\usepackage{amstext}
\usepackage{amssymb}
\usepackage[bbgreekl]{mathbbol}
\usepackage{bbm}
\usepackage{graphicx}
\usepackage{color}
\usepackage{varioref}
\usepackage{multirow,graphics}
\usepackage{epstopdf}

\newcommand{\nn}{\nonumber}

\DeclareSymbolFontAlphabet{\mathbbm}{bbold}
\DeclareSymbolFontAlphabet{\mathbb}{AMSb}

\numberwithin{equation}{section}

\def\[{\left[}
\def\]{\right]}
\def\({\left(}
\def\){\right)}
\def\<{\left<}
\def\>{\right>}
\def\d{\partial}

    \newcommand{\beq}{\begin{equation}}
    \newcommand{\eeq}{\end{equation}}
    \newcommand\beqa{\begin{eqnarray}}
    \newcommand\eeqa{\end{eqnarray}}
\newcommand\bea{\begin{array}}
\newcommand\eea{\end{array}}

\newcommand{\la}[1]{\label{#1}}
\newcommand{\eq}[1]{(\ref{#1})}

\newcommand{\pr}{\text{\textpeso}}

\definecolor{cadmiumgreen}{rgb}{0.0, 0.42, 0.24}

\makeatletter
     \@ifundefined{usebibtex}{} {}
\makeatother

\newcommand{\bra}[1]{\langle #1 |}
\newcommand{\ket}[1]{| #1 \rangle}
\newcommand{\cN}{\mathcal{N}}

\renewcommand{\<}{\langle} 
\renewcommand{\>}{\rangle} 
\renewcommand{\sl}{\mathfrak{sl}}

\newcommand{\lD}{\mathcal{D}} 
\newcommand{\lH}{\mathcal{H}}

\newcommand{\lO}{\mathcal{O}}
\newcommand{\lN}{\mathcal{N}}

\newcommand{\bB}{\textbf{B}}
\newcommand{\bC}{\textbf{C}}
\newcommand{\bbT}{\mathbb{T}}

\newcommand{\bs}{\textbf{s}}

\newcommand{\svx}{{\mathsf{x}}}

\newcommand{\svy}{{\mathsf{y}}}

\newcommand{\gl}{\mathfrak{gl}}

\newcommand{\cD}{{\cal D}}

\newcommand{\bl}{\Big(\hspace{-1.85mm}\Big(\,}
\newcommand{\br}{\,\Big)\hspace{-1.85mm}\Big)}

\newcommand{\T}{\mathbb{T}}
\newcommand{\lM}{\mathcal{M}}
\newcommand{\ee}{\mathbb{E}}
\newcommand{\lE}{\mathcal{E}}

\title{
Form-factors and complete basis of observables via separation of variables for higher rank spin chains
}

\emailAdd{nikolay.gromov$\bullet$kcl.ac.uk}
\emailAdd{nicolo.primi$\bullet$kcl.ac.uk} 
\emailAdd{paul.1.ryan$\bullet$kcl.ac.uk} 

\author[r,s]{Nikolay Gromov}

\author[r]{Nicol\`{o} Primi}

\author[r]{Paul Ryan}

\affiliation[(r)]{
Mathematics Department, King's College London,
The Strand, London WC2R 2LS, UK
}

\affiliation[(s)]{St.Petersburg INP, Gatchina, 188 300, St.Petersburg,
Russia}

\abstract{
Integrable $\sl(N)$ spin chains, which we consider in this paper, are not only the prototypical example of quantum integrable systems but also systems with a wide range of applications. For these models we use the Functional Separation of Variables (FSoV) technique with a new tool called Character Projection to compute all matrix elements of a complete set of operators, which we call {\it principal operators}, in the basis diagonalising the tower of conserved charges as determinants in Q-functions. Building up on these results we then derive similar determinant forms for the form-factors of combinations of multiple principal operators between arbitrary factorizable states, which include, in particular, off-shell Bethe vectors and Bethe vectors with arbitrary twists. We prove that the set of principal operators generates the complete spin chain Yangian. Furthermore, we derive the representation of these operators in the SoV bases allowing one to compute correlation functions with an arbitrary number of principal operators. Finally, we show that the available combinations of multiple insertions includes Sklyanin's SoV $B$ operator. As a result, we are able to \textit{derive} the $B$ operator for $\sl(N)$ spin chains using a minimal set of ingredients, namely the FSoV method and the structure of the SoV basis.

}

\begin{document}
\maketitle
\section{Introduction}

Integrability in physical systems is an extremely powerful tool often allowing one to extract exact results in very complicated systems. 
One such example is maximally supersymmetric Yang-Mills theory in $4$D ($\lN=4$ SYM) whose spectral problem, due to integrability, has been reduced to a simple set of equations on a handful of Baxter Q-functions called Quantum Spectral Curve (QSC) \cite{Gromov:2013pga}. 
At the same time, in integrable spin chains, the Q-functions serve as building blocks of the model's wave functions and correlation functions in a special 
basis called 
separation of variables (SoV) basis \cite{SklyaninFBA} leading one to believe that the same should be true in $\lN=4$ SYM. 

Remarkably, certain three point correlation functions in $\lN=4$ SYM have indeed been shown to take an incredibly simple form when expressed in terms of 
the QSC Q-functions \cite{Cavaglia:2018lxi,Giombi:2018qox,McGovern:2019sdd} and the resulting expressions are reminiscent of 
correlation functions in integrable spin chains when expressed in separated variables. This observation has been one of the main driving factors in the development 
of SoV methods for higher rank \cite{Gromov:2016itr,Maillet:2018bim,Ryan:2018fyo,Maillet:2018czd,Ryan:2020rfk,Maillet:2020ykb} integrable spin chains which was, until recently, only
applicable to the simplest rank one models with $\sl(2)$ symmetry, see \cite{Ryan:2022ybk} for a recent comprehensive review.

The operator-based SoV (OSoV) construction, going back to the original ideas of Sklyanin \cite{Sklyanin:1984sb}, has recently been supplemented with a \textit{functional} SoV (FSoV) construction \cite{Cavaglia:2019pow} allowing one to compute highly 
non-trivial quantities such as scalar products and form factors directly in separated variables bypassing the explicit operator-based construction of the SoV bases. This makes the functional approach particularly attractive in settings where an explicit 
construction of the SoV bases is complicated. Such example include infinite-dimensional systems without a highest-weight state.
For instance, FSoV has been already used to compute non-perturbative correlators directly in $\lN=4$ SYM and its cousin 4D conformal fishnet theory\footnote{The operator-based SoV construction has also been used to compute Basso-Dixon correlators in $2D$ fishnet CFT \cite{Derkachov:2018rot} and has recently seen remarkable extensions to the $4$D setting \cite{Derkachov:2019tzo,Derkachov:2020zvv,Derkachov:2021ufp}. The crucial difference with our approach is that in those papers the SoV is applied in the ``mirror" channel. This simplifies the problem of finding the SoV basis dramatically. At the same time it requires one to consider each Feynman diagram separately rather than giving a resummed non-perturbative result.} \cite{Cavaglia:2018lxi,Cavaglia:2021mft}. In 
particular, the functional SoV approach allows one to naturally compute a family of diagonal form factors $\langle 
\Psi|\partial_p \hat{I}|\Psi\rangle$, where $p$ is some parameter of the model and $\hat{I}$ is an integral of motion, using standard quantum mechanical perturbation theory 
arguments \cite{Cavaglia:2019pow,Gromov:2020fwh,Cavaglia:2021mft}. Interesting quantities which can be extracted using 
this approach are the form-factors of various \textit{local} operators, including a family of non-trivial Feynman diagrams \cite{Cavaglia:2021mft} in conformal fishnet theory. Unfortunately, it was not initially clear how to advance beyond the computation of diagonal form-factors.

In this 
paper we advance the study of correlators using the FSoV approach by tackling this issue with a novel {\it character projection} technique
and by identifying a set of $(N-1)\times (N+1)$
distinguished operators $\pr_{a,r}(u)$, which we call \textit{principal}. The operators $\pr_{a,r}(u)$ 
are polynomials in $u$, whose coefficients are certain linear combinations of elements of the spin chain monodromy matrices in anti-symmetric representations of $\sl(N)$.
In general, they do not commute with the integrals of motion, but, nevertheless,
with the help of the character projection method, we managed to compute their off-diagonal matrix elements in a simple determinant form in terms of the Q-functions.
Even more generally, we show that the same determinant form holds true  for the 
form factor $\langle \Psi_A|\pr_{a,r}(u)|\Psi_B\rangle$, where 
$|\Psi_B\rangle$ and $\langle \Psi_A|$ are two general factorisable 
states, which also includes off-shell Bethe states.

Generalising our results further, we show that the form-factors of certain anti-symmetric combinations of the principal operators also take a determinant form. Importantly, a particular case of such combinations is the SoV ${\bf B}(u)$ operator, which is at the heart of the operatorial SoV construction \cite{Sklyanin:1992sm,Smirnov2001,Gromov:2016itr,Gromov:2018cvh,Ryan:2018fyo,Ryan:2020rfk}. 
The SoV ${\bf B}$ operator has long been known as a rather mysterious object having been initially obtained by Sklyanin \cite{Sklyanin:1992sm} in analogy with SoV for classical integrable models \cite{Sklyanin:1992eu}. In higher rank systems, its relation to quantum SoV was only recently understood in \cite{Gromov:2016itr} but the precise reason for its structure remained unclear, despite its many nice properties \cite{Ryan:2018fyo,Ryan:2020rfk}. In this paper we demonstrate that ${\bf B}$ naturally follows from the interplay between the FSoV construction and the approach presented in \cite{Maillet:2018bim,Ryan:2018fyo,Ryan:2020rfk} and we derive its explicit form directly. This closes an important conceptual gap in the existing literature.

Finally, we also compute the SoV representation of all the principal operators, which allows one to construct arbitrary combinations of these operators (not only anti-symmetric). In particular we show that those operators generate the complete set of the spin chain Yangian operators $T_{ij}(u)$.
Note that at least in the finite dimensional case, this implies, via the  ``quantum inverse transform" \cite{Slavnov:2018kfx} 
that we have an access to all local symmetry generators
$\ee_{ij}^{(\alpha)}$ from which one can in turn build any physical observables in this system. We also believe this to be the case in general but we do not have a simple proof of this.

This paper is organsied as follows. In section \ref{sec:review} we review basic aspects of $\sl(N)$ spin chains and elements of the operatorial SoV construction which we will use throughout the paper. In \ref{sec3} we review the functional SoV method, which is the main tool used in this paper, and we use it to approach the computation of diagonal form-factors. In \ref{sec:sl3disc} we tackle the computation of off-diagonal correlators for the so-called \textit{principal operators} and introduce the character projection trick in the simplest but highly non-trivial setting of $\sl(3)$ spin chains. In \ref{sec5} we extend our construction to include correlators of multiple principal operators, and find the SoV $\bf B$ and $\bf C$ operators. The general $\sl(N)$ case is an almost-trivial extension, which we perform in section \ref{sec:slnextension}, which demonstrates the power of our construction.  Finally in section \ref{sec7} we prove that the principal operators form a complete basis of the observables. Four appendices supplement the main text. Appendix \ref{app:alt} contains an alternative derivation of a key relation used in the main text. Appendix \ref{dict} contains a derivation of the matrix elements of principal operators in the SoV bases. Finally, in appendix \ref{deriveB} we discuss how the FSoV method together with the SoV bases built using the ideas of \cite{Maillet:2018bim} allows one to deduce Sklyanin's SoV $B$ operator which has long been a gap in the literature. We have attached an ancillary Mathematica file with the arxiv submission of this paper which computes the SoV matrix elements of the principal operators. The file is named \verb"CodeForFormfactorsInSoVbasis.nb". 

\section{Lightning review of $\sl(N)$ spin chains and separation of variables}\label{sec:review}
Here we give a speedy review of the main notations and formulate our set-up
for the $\sl(N)$ spin chain.

\subsection{$\sl(N)$ spin chain and transfer matrices}\label{slnalg}

In order to keep our exposition short we only write the basic formulas we use throughout the paper with special emphasis on $\sl(3)$ spin chains. In this paper we consider mainly the same set-up as in \cite{Gromov:2020fwh}. 

\paragraph{$\gl(N)$ algebra.}

To define the $\sl(N)$ spin chain we introduce $L$ copies of the $\gl(N)$ algebra, one per site of the spin chain, with each copy generated by $\ee_{ij}^{(\alpha)}$ subject to the commutation relations 
\beq
\label{algebrasl}
\[\mathbb{E}^\alpha_{ij}\, ,\mathbb{E}^\beta_{kl}\]=\delta^{\alpha\beta}(\delta_{jk}\mathbb{E}^\alpha_{il}-\delta_{li}\mathbb{E}^\alpha_{kj})\,.
\eeq
We will consider a spin-$\bs$ highest-weight representation of this algebra with highest-weight state $|0\rangle$ satisfying 
\begin{equation}
\begin{split}
    & \mathbb{E}^\alpha_{ij} |0\rangle = 0,\quad i<j \\
    & \mathbb{E}^\alpha_{ii} |0\rangle = \omega_{i}|0\rangle \,,
\end{split}
\end{equation}
where $\omega_1=-\bs$ and $\omega_i=+\bs$ for $i>2$. This is the simplest non-compact representation which can be considered and we have chosen it for simplicity to illustrate our main results, but we believe all the main statements can be easily extended to more general representations.

One can build the representation space of one site $\alpha$ of the spin chain as polynomials of $N-1$ variables $z^\alpha_1, \dots z^\alpha_{N-1}$. The full representation space will just be the tensor product of the representations at each site, for a total of $L(N-1)$ degrees of freedom. The explicit form of the generators can be found in \cite{Gromov:2020fwh} for $N=2$ and $N=3$ and the general $N$ construction can be found  for example in \cite{Derkachov:2006fw,antonenko2021gelfand}.

Using the $\gl(N)$ algebra we define a Lax operator acting on the $\alpha$-th site
\beq
\label{Laxop}
\mathcal{L}^{(\alpha)}_{ij}(u)=u\,\delta_{ij}+i\,\mathbb{E}^\alpha_{ji}\,.
\eeq
As usual, the $\alpha$-th Lax operator will act on the tensor product of a quantum space, i.e. the representation space of the $\alpha$-th site of the spin chain, and an auxiliary space $\mathbb{C}^N$. We can then build the monodromy matrix by taking a product in the auxiliary space of the Lax operators at every site
\begin{equation}\label{transfer11}
    T_{ij}(u)=\sum_{k_1}\dots\sum_{k_{L-1}} \mathcal{L}_{i k_1}^{(1)}\left(u-\theta_{1}\right) \mathcal{L}_{k_{1} k_{2}}^{(2)}\left(u-\theta_{2}\right) \ldots \mathcal{L}_{k_{L-1} j}^{(L)}\left(u-\theta_{L}\right) 
\end{equation}
where $\theta_\alpha$ denote the spin chain inhomogeneities which we take to be real. The monodromy matrix satisfies the following commutation relation, known as RTT relations 
\begin{equation}\label{yangiangens}
    -i(u-v)[T_{jk}(u),T_{lm}(v)] = T_{lk}(v)T_{jm}(u)-T_{lk}(u)T_{jm}(v) \,,
\end{equation}
which defines the Yangian algebra $\mathcal{Y}(\gl(N))$ with generators $T_{ij}(u)$.

The fundamental transfer matrix $\T(u)$ which generates integrals of motion is defined to be the trace of the monodromy matrix
\begin{equation}
    \T(u)={\rm tr}\left(T(u)\right)=T_{11}(u) + \dots + T_{NN}(u)
\end{equation}
and satisfies $[\T(u),\T(v)]=0$. Taking the trace, however, will result in integrals of motion which have degenerate spectrum (due to the preserved $\sl(n)$ symmetry). To lift the degeneracies we introduce a twist $G$ in the monodromy matrix and instead define 
\begin{equation}
 \bbT(u) = {\rm tr}(T(u)G)
\end{equation}
where $G$ is an $N\times N$ matrix. The twisting preserves the commutativity $[\T(u),\T(v)]=0$, guaranteeing that the twisted transfer matrix continues to produce commuting integrals of motion.
For a diagonalisable $G$, without loss of generality we can choose $G$ to be diagonal with distinct eigenvalues $\lambda_j$, $j=1,\dots,N$. As we will see, for our purposes it is much more convenient, by adjusting the frame with a suitable $\sl(N)$ rotation, to choose $G$ to be the so-called companion matrix with entries 
\begin{equation}\label{companion}
    G_{ij}=(-1)^{j-1}\chi_j\delta_{i1}+\delta_{i,j+1}\,,
\end{equation}
where $\chi_j$ are the elementary symmetric polynomials in the eigenvalues $\lambda_i$
\begin{equation}
    \prod_{j=1}^N (t+\lambda_j) =\displaystyle \sum_{r=0}^N t^{N-r}\chi_r\,.
\end{equation}
Note that $\chi_r$ are the characters of the anti-symmetric representations of $GL(N)$. The usefulness of this choice of twist in the SoV framework has now been extensively demonstrated \cite{Ryan:2018fyo,Ryan:2020rfk,Gromov:2020fwh}. Most notably, the separated variable bases producing factorised wave functions for the integrals of motion are independent of the twist eigenvalues in this frame. We will return to this point later.

From the definition of the Lax operator~\eqref{Laxop} we see that the transfer matrix is a polynomial in the spectral parameter $u$ of degree $L$. Therefore we can think of the coefficients of the powers of $u$ as integrals of motion (IoM). Since there are only $L$ independent IoMs in $\mathbb{T}$, we are still missing $L(N-2)$ IoMs to grant integrability\footnote{The Hilbert space of the spin chain is the space of polynomials in $L(N-1)$ variables.}. Hence we need to introduce additional transfer matrices in anti-symmetric representations of $\sl(N)$, denoted as $\mathbb{T}_{a}(u)$ \footnote{Higher transfer matrices are usually denoted $\T_{a,s}(u)$ with $s=1$ for anti-symmetric representations. Since we will not use any other transfer matrices we simply denote $\T_{a,1}=\T_{a}$.} with $\T(u)=\T_1(u)$. These are easily obtained by the fusion procedure \cite{Zabrodin:1996vm} for the Lax operator, where we take anti-symmetric products of the fundamental one \eqref{Laxop} with shifts in the spectral parameter as:
\beq
\label{Laxfusion}
\mathcal{L}_{\bar{j}, \bar{k}}^{(a)}=\mathcal{L}^{j_{1}}{ }_{\left[k_{1}\right.}\left(u+i \frac{a-1}{2}\right) \mathcal{L}_{k_{2}}^{j_{2}}\left(u+i \frac{a-3}{2}\right) \ldots \mathcal{L}_{\left.k_{a}\right]}^{j_{a}}\left(u-i \frac{a-1}{2}\right)\,,
\eeq
where $\bar{j}=\{j_1,\dots,j_L\}$ and $\bar{k}=\{k_1,\dots,k_L\}$.

We also need to fuse the twist matrix in a similar way, obtaining:
\beq\label{fusedtwist}
G_{\bar{j}, \bar{k}}^{(a)}=G_{[k_{1}}^{j_{1}} G_{k_{2}}^{j_{2}} \ldots G_{k_{a}]}^{j_{a}}\,.
\eeq
Note that although the lower indices are explicitly anti-symmetrised the anti-symmetrisation is also present in the upper indices.

A feature of the fused twist matrix which will play a very important role later in the text is the following -- when $G$ is the companion twist matrix \eqref{companion} then all fused twists $G^{(a)}$ are linear in characters $\chi_r$, $r=0,\dots,N$ with $\chi_0=1$. Clearly this is true for $G$ itself but the fact that it holds for all $G^{(a)}$ is a bit surprising as these are degree $a$ polynomials in the entries of $G$. To prove this consider \eqref{companion}: the $\delta_{i1}$ ensures that $\chi_r$ can only appear in \eqref{fusedtwist} when an index in the set $j_1,\dots,j_a$ is equal to $1$. Any term which is at least quadratic in characters would require at least two such terms in this set to be $1$, but such a term would then vanish due to the anti-symmetry of these indices. It is also easy to verify that for any $a=1,\dots,N-1$ each character $\chi_r$ appears with non-zero coefficient in the twist matrix.

Finally, the transfer matrix in the antisymmetric representation $\mathbb{T}_{a}$ is obtained in the same way as the fundamental one:
\beq\label{highertransfer}
\mathbb{T}_{a}(u)=\sum_{\bar{b}, \bar{b}_{i}} \mathcal{L}_{\bar{b} \bar{b}_{1}}^{a(1)}\left(u-\theta_{1}\right) \mathcal{L}_{\bar{b}_{1} \bar{b}_{2}}^{a(2)}\left(u-\theta_{2}\right) \ldots \mathcal{L}_{\bar{b}_{L-1} \bar{b}_{L}}^{a(L)}\left(u-\theta_{L}\right) G_{\bar{b}_{L} \bar{b}}^{(a)}\,.
\eeq
The transfer matrices generate a mutually commuting family of integrals of motion 
\begin{equation}
    [\T_{a}(u),\T_b(v)]=0\,.
\end{equation}
Since $\mathcal{L}^a$ is a polynomial in $u$ of degree $a$, the corresponding transfer matrix will have degree $a L$. Since in principle we can have $a=1\,\dots,N$, it looks like we might have too many IoMs. However, it turns out that there are some trivial prefactors of $u$ in $\mathbb{T}_{a}$ for $a>1$. For example, the so-called quantum determinant $\mathbb{T}_{N}$ is completely non-dynamical i.e. it is proportional to the identity operator. By removing all such scalar prefactors the family of transfer matrices $\mathbb{T}_{a}$, $a={1,\dots,N-1}$ contains precisely $L(N-1)$ nontrivial IoMs, matching the numbers of dofs and guaranteeing integrability of our spin chain \footnote{Such naive counting should be handled with care as the number of degrees of freedom is not a well-defined notion for quantum systems. The number of independent integrals of motion may also drop for certain representations, see \cite{Maillet:2018bim}. Note that none of the details of this paper rely on such subtleties.}. We can introduce reduced transfer matrices $t_{a}(u)$, which are polynomials of degree $L$ related to the original transfer matrices as
\begin{equation}\label{transfertrivial}
    \T_{a}(u) = t_{a} \left(u+\frac{i}{2}(a-1) \right)\prod_{k=1}^{a-1}Q_\theta^{[2\bs-2k+a-1]}(u)\,,
\end{equation}
where we introduced the polynomial $Q_\theta(u)=\prod_{\alpha=1}^L(u-\theta_\alpha)$ and use the standard notation for shifts of the spectral parameter 
\begin{equation}
    f^{[n]}(u):=f\left(u+\frac{i}{2}n\right),\quad f^\pm(u) =f(u\pm\tfrac{i}{2}), \quad f^{\pm\pm}(u) =f(u\pm i)\,.
\end{equation}
We can expand $t_{a}$ into a family of mutually commuting integrals of motion $\hat{I}_{a,\beta}$
\begin{equation}\label{integralsofmotion}
    t_{a}(u) = \chi_a\, u^L+\displaystyle \sum_{\beta=1}^{L} u^{\beta-1}\hat{I}_{a,\beta}\,.
\end{equation}
We will define the right (left) eigenstates of the mutually commuting transfer matrices as $|\Psi\rangle$ ($\langle\Psi|$), and the relative eigenvalue of $t_{a}$ as $\tau_{a}$, so that:
\beq
t_{a}(u)|\Psi\rangle=\tau_a(u)|\Psi\rangle,\quad \langle\Psi| t_a(u)=\langle\Psi| \tau_a(u)\,.
\eeq
Similarly for the integrals of motion $\hat{I}_{a,\beta}$ we have
\beq
\hat{I}_{a,\beta}|\Psi\rangle=I_{a,\beta}|\Psi\rangle,\quad \langle\Psi| \hat{I}_{a,\beta}=I_{a,\beta}\langle\Psi|\;.
\eeq

\subsection{Principal operators}
\label{principalop}
A major goal in this paper will be to compute the matrix elements of (sums of) certain monodromy matrix entries between two transfer matrix eigenstates and their generalisation to arbitrary factorisable states. We will refer to these particular monodromy matrix entries as \textit{principal operators}.

The principal operators are defined as follows. As was demonstrated above, each of the fused companion twist matrices $G^{(a)}$ are linear in the characters $\chi_r$. As such, each of the transfer matrices $t_a(u)$ admits an expansion\footnote{In \LaTeX $\, $one can use the \textbackslash textpeso command to generate this symbol.}
\begin{equation}
\label{genfunc}
    t_a(u) \equiv \displaystyle \sum_{r=0}^N \chi_r\, \pr_{a,r}(u)\,.
\end{equation}
We call the operators $\pr_{a,r}(u)$ principal and the reason for their importance will become clear in section \ref{sec:sl3disc}. Note that they are independent of the twist eigenvalues $\lambda_j$ as all twist dependence of the transfer matrices is contained in the characters $\chi_r$.

For example, the transfer matrix $t_1(u)$ can be expanded as 
\begin{equation}
    t_1(u) = \displaystyle \sum_{j=1}^{N-1}\chi_0 T_{j,j+1}(u) + \displaystyle \sum_{r=1}^{N} \chi_r (-1)^{r-1}T_{r1}(u)\,,
\end{equation}
where $\chi_0=1$.

Similar expansions can be performed for the higher transfer matrices $t_{a}(u)$. For this we need to introduce \textit{quantum minors} defined by 
\begin{equation}
    T\left[^{i_1\dots i_a}_{j_1\dots j_a}\right](u) = \displaystyle \sum_{\sigma} (-1)^{|\sigma|} T_{i_1 j_{\sigma(1)}}^{[a-1]}(u)\dots T_{i_a j_{\sigma(a)}}^{[-a+1]}(u)
\end{equation}
where the sum is over all elements $\sigma$ of the permutation group on $a$ indices. 
The transfer matrices $\T_a(u)$ in the $a$-th antisymmetric representation are then given by 
\begin{equation}
    \T_a(u) = \displaystyle \sum_{1\leq i_1 < \dots< i_a \leq N} T\left[^{i_1\dots i_a}_{j_1\dots j_a}\right](u)G_{j_1 i_1}\dots G_{j_a i_a}\,.
\end{equation}
As a result of the summation condition $1\leq i_1 < \dots < i_a \leq N$ the coefficient of each $\chi_r$ is a sum of quantum minors with distinct upper indices which cannot cancel each other and as a result the coefficient of each $\chi_r$ is non-zero as long as $1\leq a \leq N-1$.

While most principal operators are given by large sums over quantum minors things simplify for $a=N-1$ as the $N-1$-th anti-symmetric representation monodromy matrix is simply equal to the quantum-inverse matrix of $T(u)$ divided by a trivial factor. We introduce the notation $T^{ij}$ for these operators, defined by 
\begin{equation}\label{monupup}
    T^{ij}(u)\, \prod_{k=1}^{N-1}Q_\theta^{[2(\bs-k)]}(u)= T\left[^{1\, \dots\, \hat{j}\,\dots\, N}_{1\, \dots\, \hat{i}\,\dots\, N} \right]\left(u-\frac{i}{2}(N-2) \right)
\end{equation}
where the notation $\hat{i}$, $\hat{j}$ means that the corresponding index is removed. It is then easy to derive 
\begin{equation}
    t_{N-1}(u) = \displaystyle \sum_{r=0}^{N-1}\chi_r\, T^{r+1,N}(u) - \chi_N \displaystyle \sum_{j=1}^{N-1} T^{j+1,j}(u)\,.
\end{equation}

We will write out explicitly the principal operators in terms of monodromy matrix elements $T_{ij}$ for the special cases of $\sl(2)$ and $\sl(3)$. \paragraph{$\sl(2)$ case.} 
In this case we have 
\begin{equation}
    t_1(u) = T_{12}(u) + \chi_1 T_{11}(u) - \chi_2 T_{21}(u)
\end{equation}
and hence 
\begin{equation}\label{sl2principal}
    \pr_{1,0}(u) =T_{12}(u),\quad \pr_{1,1}(u)  =T_{11}(u),\quad \pr_{1,2}(u) = -T_{21}(u)\,.
\end{equation}

\paragraph{$\sl(3)$ case.} For the special case of $\sl(3)$ there are only two non-trivial transfer matrices $t_1(u)$ and $t_2(u)$ which in the notations described above admit the expansions of table~\ref{table}, where $t_2$ is written both in terms of the original monodromy elements $T_{ij}$ and the elements $T^{ij}$ defined in~\eqref{monupup}
\renewcommand{\arraystretch}{1.7}
\begin{center}\label{table}
\begin{tabular}{| c | l | l |}
\hline
 $\pr_{1,0}(u)=$ & $+T_{12}+T_{23} $ &\\
 $\pr_{1,1}(u)=$ & $+T_{11} $ &\\
 $\pr_{1,2}(u)=$ & $-T_{21} $ &\\
 $\pr_{1,3}(u)=$ & $+T_{31} $ &\\
 \hline
 \hline
 $\pr_{2,0}(u)=$ 
 & $\left(T_{12}T_{23}^{--}-T_{13}T_{22}^{--}\right)/Q_\theta^{[2\bs-2]}$
 & $+T^{13}/Q_\theta^{[2\bs-2]}$\\
 $\pr_{2,1}(u)=$ 
 & $\left(T_{11}T_{23}^{--}-T_{13}T_{21}^{--}\right)/Q_\theta^{[2\bs-2]}$
 & $+T^{23}/Q_\theta^{[2\bs-2]}$ \\
 $\pr_{2,2}(u)=$ 
 & $\left(T_{11}T_{22}^{--}-T_{12}T_{21}^{--}\right)/Q_\theta^{[2\bs-2]}$& $+T^{33}/Q_\theta^{[2\bs-2]}$\\
 $\pr_{2,3}(u)=$ 
 & $
 \left(-T_{11}T_{32}^{--}+T_{12}T_{31}^{--}-T_{21}T_{33}^{--}+T_{23}T_{31}^{--}\right)/Q_\theta^{[2\bs-2]}
 $& $-(T^{21}+T^{32})/Q_\theta^{[2\bs-2]}$\\
 \hline
\end{tabular}
\end{center}

Since the transfer matrices $t_a(u)$ admit the expansion \eqref{integralsofmotion} into integrals of motion $\hat{I}_{a,\alpha}$ it clearly follows that each $\hat{I}_{a,\alpha}$ also admits a linear expansion into characters $\chi_r$. We will denote the coefficients of the characters in this expansion $I_{a,\alpha}^{(r)}$ and so
\beq\la{Ichi}
\hat I_{a,\alpha}=\sum_{r=0}^N
\chi_r\, \hat I_{a,\alpha}^{(r)}\;.
\eeq

Finally, since the transfer matrices commute for different values of the spectral parameters $[t_a(u),t_b(v)]=0$ we see that by expanding into principal operators we obtain the relation
\begin{equation}
\la{commP}
    \sum_{r,s}\chi_r \chi_s [\pr_{a,r}(u),\pr_{b,s}(v)]=0\,.
\end{equation}
As this should hold for arbitrary twists $\lambda$ it is easy to see\footnote{For example one can change variables from $\lambda_i,\;i=1,\dots,N$ to $\chi_i,\;i=1,\dots,N$. The Jacobian of such transformation is simply a Vandermonde determinant of $\lambda$'s so this is always possible for generic $\lambda$'s.
After that \eq{commP} becomes a quadratic polynomial in $N$ independent variable $\chi_i,\;i=1,\dots,N$ which is identically zero, which is only possible if all coefficients vanish.
} 
that the above expression implies
$[\pr_{a,r}(u),\pr_{b,s}(v)]+[\pr_{a,s}(u),\pr_{b,r}(v)]=0$ which in particular gives $[\pr_{a,r}(u),\pr_{b,r}(v)]=0$, that is principal operators corresponding to the same character index $r$ form a commutative family.

\subsection{Baxter equations}

The spectrum of transfer matrices can be determined by means of the Baxter equations. These are finite-difference equations for functions denoted $Q_i$ and $Q^i$, $i=1,\dots,N$ called Q-functions.
The Baxter equations can be conveniently written as 
\begin{equation}\la{bax}
    \lO Q_i =0,\quad \lO^\dagger Q^i=0\;.
\end{equation}
We refer to $\lO$ as the Baxter operator and $\lO^\dagger$ as the dual Baxter operator.
They are finite-difference operators defined as
\begin{equation}\label{dualbax}
    \lO^\dagger = \sum_{a=0}^N (-1)^a \tau_a(u)\lD^{N-2a}\;\;,\;\;
    \lO = \sum_{a=0}^N (-1)^a \lD^{2a-N}\tau_a(u)\varepsilon(u)
\end{equation}
where $\lD$ is the shift operator satisfying $\lD\,f(u)=f(u+\frac{i}{2})$, $\tau_a,\,\,a=1,\dots,N-1$ are the eigenvalues of the reduced transfer matrices $t_a$ and we have denoted:
\begin{equation}
    \tau_0(u) = Q_\theta^{[2\bs]},\quad \tau_N(u) = \chi_N Q_\theta^{[-2\bs]},\quad Q_{\theta}(u)=\prod_{\alpha=1}^L(u-\theta_{\alpha})\,,
\end{equation}
and finally $\varepsilon(u)$ is the function 
\begin{equation}
    \varepsilon(u) = \prod_{\beta=1}^L \frac{\Gamma(\bs-i(u-\theta_\beta))}{\Gamma(1-\bs-i(u-\theta_\beta))}\,.
\end{equation}
The Q-functions can be characterised by their large-$u$ asymptotics which are related to the twist eigenvalues $\lambda_j$. We label the Q-functions so that their asymptotics read 
\begin{equation}
    Q_j \sim \lambda_j^{iu}u^{M_i},\quad Q^j \sim \lambda_j^{-iu}u^{M^i}
\end{equation}
where $M_i$ and $M^i$ are some real quantum numbers determined by the state in question. It is also possible to choose the Q-functions so that one of the Q-functions $Q_i$, which we take to be $Q_1$, and $N-1$ of the Q-functions $Q^i$, which we take to be $Q^{1+a}$, $a=1,\dots,N-1$, are \textit{twisted polynomials}, meaning that they have the structure 
\begin{equation}
    Q_1(u) = \lambda_1^{iu}q_1(u),\quad Q^{1+a}(u) = \lambda_{1+a}^{-iu} q^{1+a}(u)\,,
\end{equation}
where $q_1(u)$ and $q^{1+a}(u)$ are polynomials. Requiring that the equations \eq{bax} have twisted-polynomials solutions $Q_1$ and $Q^{2},\dots,Q^N$ we constrain the possible values for the coefficients $\tau_a(u)$, which gives the spectrum of the IoM for the spin chain in this approach \cite{Krichever:1996qd}.

\subsection{SoV bases and wave functions} 
The SoV bases for the representations considered in this work were constructed in \cite{Gromov:2020fwh}. The left SoV basis $\bra{\svx}$ is obtained by diagonalising the $\bB$ operator \cite{Sklyanin:1992sm}\cite{Smirnov2001}\cite{Gromov:2016itr}
and the right SoV basis $|\svy\rangle$ is obtained by diagonalising the $\bC$ operator \cite{Gromov:2019wmz,Gromov:2020fwh}, see \cite{Gromov:2020fwh} for definitions of the $\bB$ and $\bC$ operators for the $\sl(N)$ case. For the specific case of $\sl(3)$ we have
\beqa
{\bf B}(u)&=&
-T_{11}(T_{11}^{--}T_{22}-T_{21}^{--}T_{21})-
T_{21}(T_{11}^{--}T_{23}-T_{21}^{--}T_{13})\;,\\
{\bf C}(u)&=&
-T_{11}(T_{11}T_{22}^{++}-T_{21}T_{21}^{++})-
T_{21}(T_{11}T_{23}^{++}-T_{21}T_{13}^{++})\;.
\eeqa
Using the RTT relations \eqref{yangiangens} it is possible to rewrite these expressions in a slightly different form:
\beqa\label{BandC}
{\bf B}(u)&=&
-T_{11}(T_{11}T_{22}^{--}-T_{22}T^{--}_{21})-
(T_{11}T_{23}^{--}-T_{13}T_{21}^{--})T_{21}\;,\\
{\bf C}(u)&=&
-T_{11}(T_{11}^{++}T_{22}-T^{++}_{22}T_{21})-
(T^{++}_{11}T_{23}-T^{++}_{13}T_{21})T_{21}\;.
\eeqa
This simple rewriting allows us to express the ${\bf B}$ and ${\bf C}$ operators in terms of the principal operators (after removing the trivial non-dynamical factor) in an ordering which will be convenient later
\beq\la{BCinP}
-\frac{{\bf B}(u)}{Q_\theta^{[2\bs-2]}}\equiv
{{\bf b}(u)}
=\pr_{1,1}\pr_{2,2}-\pr_{2,1}\pr_{1,2}\;\;,\;\;
-\frac{{\bf C}(u)}{Q_\theta^{[2\bs]}}
\equiv
{{\bf c}(u)}
=
\pr_{1,1}\pr^{++}_{2,2}-\pr^{++}_{2,1}\pr_{1,2}\;.
\eeq
The spectrum of $\bB(u)$ was first found in \cite{Gromov:2016itr}
and then generalised for general representations in \cite{Ryan:2018fyo}.
In our case we get
\beq
\langle \svx|{\bf b}(u) = \langle \svx|\prod_{\alpha=1}^L\prod_{a=1}^{N-1}(u-\svx_{\alpha,a})\;\;,\;\;
{\bf c}(u)|\svy\rangle = \prod_{\alpha=1}^L\prod_{a=1}^{N-1}(u-\svy_{\alpha,a})|\svy\rangle
\eeq
where each SoV basis element $\langle \svx|$ and $|\svy\rangle$ is parameterised by $L(N-1)$ numbers $\svx_{\alpha,a}$ and $\svy_{\alpha,a}$ respectively, with $\alpha=1,\dots, L$ and $a=1,\dots,N-1$, which are of the form 
\begin{equation}\label{svxsvy}
    \svx_{\alpha,a}=\theta_\alpha + i(\bs+n_{\alpha,a}),\quad \svy_{\alpha,a}=\theta_\alpha + i(\bs+m_{\alpha,a}+1-a)\,,
\end{equation}
where $n_{\alpha,a}$ and $m_{\alpha,a}$ are non-negative integers subject to the constraints $n_{\alpha,1}\geq \dots \geq n_{\alpha,N-1}\geq 0 $ and $m_{\alpha,1}\geq \dots \geq m_{\alpha,N-1}\geq 0$ with each possible configuration corresponding to a basis state. In the polynomial representation described above in section \ref{slnalg} the SoV ground states $\langle 0|$ and $|0\rangle$ (with all $n$'s or $m$'s being zero) can be shown to be constant polynomials.
It is convenient to fix their normalization to be
\beq\la{xy0}
\langle 0|=1\;\;,\;\;|0 \rangle = 1\;.
\eeq
\paragraph{SoV charge.}
A useful object proposed in \cite{Gromov:2019wmz} is the so-called SoV charge operator ${\bf N}$. It commutes with the ${\bf B}(u)$ and ${\bf C}(u)$ operators and is diagonalised in both SoV bases $|\svy\rangle$ and $\langle \svx|$ and counts the number of ``excitations" above the SoV ground state. More precisely:
\begin{equation}\label{SoVcharge}
    {\bf N}|\svy\rangle = \left(\displaystyle \sum_{\alpha,a}m_{\alpha,a}\right)|\svy\rangle ,\quad \langle\svx| {\bf N}= \langle\svx|\left(\displaystyle \sum_{\alpha,a}n_{\alpha,a}\right)\,.
\end{equation}
It can be obtained as the first non-trivial coefficient in the large $u$ expansion of $\bB(u)$ or $\bC(u)$.

\paragraph{Wave function factorisation.}
The $\langle \svx|$ basis factorises the wave functions $\Psi(\svx)$ of the right transfer matrix eigenstates $|\Psi\rangle$ whereas the basis $|\svy\rangle$ factorises the wave functions $\Psi(\svy)$ of the left transfer matrix eigenstates $\langle\Psi|$.
The right wave functions (i.e. eigenfunctions of the transfer matrices) are then given explicitly by
\begin{equation}\label{wave1}
    \Psi(\svx):=\langle \svx|\Psi\rangle = \displaystyle \prod_{\alpha=1}^L \prod_{a=1}^{N-1} Q_1(\svx_{\alpha,a})
\end{equation}
and the left wave functions are given by
\begin{equation}\label{wave2}
    \Psi(\svy):=\langle \Psi|\svy\rangle = \displaystyle \prod_{\alpha=1}^L \prod_{a=1}^{N-1} \det_{1\leq a,b\leq N-1}Q^{a+1}\left(\svy_{\alpha,b}+\frac{i}{2}(N-2)\right)\;.
\end{equation}
For the above to be correct one should of course fix the normalisation of $\Psi$, for example by fixing $\langle 0|\Psi\rangle $ and $\langle \Psi|0\rangle $ in agreement with \eq{wave1} and \eq{wave2}. After that \eq{wave1} and \eq{wave2} stay true for any element of the SoV basis \cite{Gromov:2016itr,Ryan:2018fyo,Ryan:2020rfk,Gromov:2020fwh}.

The scalar product between two states, normalised as described above, is then given by 
\begin{equation}\label{SoVscalarprod}
    \langle\Psi_A |\Psi_B\rangle = \displaystyle \sum_{\svx,\svy}  \Psi_A(\svy)\mathcal{M}_{\svy,\svx}\Psi_B(\svx)\,.
\end{equation}
Here $\mathcal{M}_{\svy,\svx}$ is the measure in the SoV basis. 
It can be also written in terms of the dual bases $|\svx\rangle$
and $\langle\svy|$, which are defined such that $\langle\svx|\svx'\rangle=\delta_{\svx,\svx'}$
and $\langle\svy|\svy'\rangle=\delta_{\svy,\svy'}$, as
the overlap $\mathcal{M}_{\svy,\svx} = \langle\svy|\svx\rangle$.
In general the overlaps $\langle\svy|\svx\rangle$ are not diagonal and so the matrix $\mathcal{M}_{\svy,\svx}$ could be potentially quite complicated. Nevertheless it is known explicitly from \cite{Gromov:2020fwh}, and we review its structure next.

\subsection{SoV measure}

The explicit form of the measure, worked out in \cite{Gromov:2020fwh}, is given by \footnote{There is a typo in \cite{Gromov:2020fwh} where the sign factor $s_{\bf L}$ does not appear. However, it is correctly included in the Mathematica code contained in that paper.} 
\beq\label{measure}
\lM_{\svy,\svx}=
s_{\bf L}\sum_{k}
\left.
{\rm sign}(\sigma)
\(\prod_{a=1}^{N-1}
\frac{
\Delta_a
}{
\Delta_\theta
}\)
\prod_{\alpha=1}^L\prod_{a=1}^{N-1} \frac{r_{\alpha,n_{\alpha,a}}}{r_{\alpha,0}}\right|_{\sigma_{\alpha,a}=k_{\alpha,a}-m_{\alpha,a}+a}.
\eeq
See Appendix \ref{dict} for a detailed derivation of this formula. We will now summarise the notations we use, following \cite{Gromov:2020fwh}. $s_{\bf L}$ is a simple sign factor 
\begin{equation}
    s_{\bf L}=(-1)^{\frac{L}{4}(L-1)(N^2+N-2)}\,.
\end{equation}
$\sigma$ denotes a permutation of $L$ copies of the numbers $\{1,2,\dots,N-1\}$
\begin{equation}\la{ini}
    \{\underbrace{1,\dots,1}_{L},\dots,\underbrace{N-1,\dots,N-1}_{L}\}
\end{equation}
with $\sigma_{\alpha,a}$ denoting the number at position $a+(\alpha-1)(N-1)$. $\sigma^0$ denotes the identity permutation on this set and so $\sigma^0_{\alpha,a}=a$. The signature of the permutation ${\rm sign}(\sigma)$ is $\pm 1 $ depending on the number of elementary permutations needed to bring the ordered set $u_{\sigma^{-1}(1)}\cup u_{\sigma^{-1}(2)}\dots \cup
u_{\sigma^{-1}(N-1)}$ to the canonical order $u_{1,1},u_{1,2},\dots,u_{L,N-1}$ where $u_{\sigma^{-1}(a)}=\{u_{\alpha,b}: \sigma_{\alpha,b}=a\}$. Whereas ${\rm sign}(\sigma)$
could be ambiguous due to different possible orderings inside $\sigma^{-1}(a)$, the combination with the Vandermondes $\Delta_b$ is well defined. There are  $\frac{(N-1)L!}{L!^{N-1}}$ possible permutations $\sigma$, and if $\sigma$ is not such a permutation we define ${\rm sign}(\sigma)=0$.

Since the SoV charge operator~\eqref{SoVcharge}  commutes with both ${\bf b}(u)$ and ${\bf c}(u)$, $\lM_{\svy,\svx}$ is only non-zero if the states $\langle\svx|$ and $|\svy\rangle$ have the same SoV charge eigenvalue. Furthermore, $\lM_{\svy,\svx}$ is only non-zero if there exists a permutation $\sigma$ of the number \eq{ini} such that
\begin{equation}\label{measperms}
    m_{\alpha,a}=n_{\alpha,a}-\sigma_{\alpha,a}+a
\end{equation}
for each $\alpha,a$. There are distinct dual basis states $|\svx\rangle$ with the same value of $n_{\alpha,a}$ and hence there are multiple permutations satisfying \eqref{measperms}. We denote such inequivalent permutations (within each $\alpha)$ by $k$ which we then sum over. The sum over $k$ is needed only in a limited number of cases, for example in the $\sl(3)$ case only $k=n$ is possible.

In~\eqref{measure}, $\Delta_b$, which depends on $\sigma$, denotes the Vandermonde determinant constructed from all $\svx_{\alpha,a}$ for which $\sigma_{\alpha,a}=b$ and $\Delta_\theta$ denotes the Vandermonde determinant built from $\theta$'s
\begin{equation}
    \Delta_\theta=\displaystyle \prod_{\alpha<\beta}(\theta_\alpha-\theta_\beta)\,.
\end{equation}
Finally, the function $r_{\alpha,n}$ is defined as 
\beq \label{resmu}r_{\alpha, n}=-\frac{1}{2 \pi} \prod_{\beta=1}^{L}\left(n+1-i \theta_{\alpha}+i \theta_{\beta}\right)_{2 \mathbf{s}-1}\,,
\eeq
where $(z)_s=\frac{\Gamma(s+z)}{\Gamma(z)}$ is the Pochhammer symbol.

An explicit Mathematica implementation of the measure is provided in \cite{Gromov:2020fwh}. Although the scalar product can be expressed as the sum \eqref{SoVscalarprod} it is most conveniently expressed using the functional SoV (FSoV) formalism which we now review. 

\section{Functional Separation of Variables method}\la{sec3}
In this section we review the key idea of the functional separation of variables method of~\cite{Cavaglia:2019pow}. We will then extend this method in section \ref{sec:sl3disc} by introducing the character projection tool.

\subsection{Functional orthogonality and scalar product}

The key relation in the functional SoV approach is the adjointness condition \cite{Cavaglia:2019pow,Gromov:2019wmz,Gromov:2020fwh}
\begin{equation}\label{adjointness}
    \bl f \lO^\dagger g \br_\alpha = \bl g M_\alpha \lO f \br_\alpha \,,
\end{equation}
where the bracket $\bl f(w) \br_\alpha$ is defined by 
\begin{equation}
\bl f(w) \br_\alpha = \int^\infty_{-\infty} {\rm d}w\,\mu_\alpha(w) f(w)\,,
\end{equation}
the measure factor $\mu_\alpha$ is given by \cite{Gromov:2020fwh}
\begin{equation}
\la{measure}
    \mu_\alpha(w) = \frac{1}{1-e^{2\pi(w-\theta_\alpha-i\bs)}}\displaystyle \prod_{\beta=1}^L \frac{\Gamma(\bs-i(w-\theta_\beta))}{\Gamma(1-\bs-i(w-\theta_\beta))}\,,
\end{equation}
and $M_\alpha$ is some irrelevant factor which does not depend on the functions $f$ and $g$. By appropriate re-definitions of the Baxter equations it is possible to set $M_\alpha=1$, but it is nontrivial in our current conventions.

We will be interested in particular in the case where the functions $f$ and $g$ are certain Q-functions $Q_1$ and $Q^2,\dots,Q^N$ or functions with similar analytic properties.
The way to compute these integrals is to close the contour in the upper half plane and write them as a sum of residues. However, we need to ensure that the integrals actually converge and that the contour can be closed in this way without changing the result. In order to do so, it is sufficient to impose constraints on the twists that we find inside the Q-functions, as in \cite{Gromov:2020fwh}, which read
\begin{equation}
    0<{\rm arg}\lambda_a-{\rm arg}\lambda_1<2\pi,\quad a=2,\dots,N\,.
\end{equation}
Once we do this, 
we can replace the integral by the sum of the residues in the upper half-plane. Since the Q-functions are analytic everywhere, the only contribution comes from the simple poles of the measure factor \eqref{measure}. These poles are situated at $w=\theta_{\alpha}+i\bs+i n,\,n\in\mathbb{Z}_{\geq 0}$. As such we can write the bracket as an infinite sum of the residues at the poles of the measure:
\beq\la{brsum}
\bl 
f(w)
\br_\alpha =\sum_{n=0}^\infty \frac{r_{\alpha,n}}{r_{\alpha,0}} f(\theta_{\alpha}+i\bs+i n)\;,
\eeq
with $r_{\alpha,n}$ being the residue of $\mu_{\alpha}$ at the pole $\theta_{\alpha}+i \bs+i n$:
\beq
r_{\alpha,n}=-\frac{1}{2\pi}\prod_{\beta=1}^L (n+1-i\theta_\alpha+i\theta_\beta)_{2\bs-1}\;,
\eeq
where $(z)_s=\frac{\Gamma (s+z)}{\Gamma (z)}$ denotes the Pochhammer symbol and we have included the overall normalisation $r_{\alpha,0}$ for convenience. 

\subsection{Basic idea of Functional SoV}

To demonstrate the basic idea of the FSoV notice that the adjointness condition \eqref{adjointness} implies in particular
\begin{equation}
    \bl f \lO^\dagger Q^{1+a}\br_\alpha=0= \bl Q_1 \lO^\dagger g\br_\alpha=0,\quad \alpha=1,\dots,L,\quad a=1,\dots,N-1
\end{equation}
and so if we pick $Q^{1+a}_A$ and $Q_1^B$ to be the Q-functions associated to two transfer matrix eigenstates $\ket{\Psi_A}$ and $\ket{\Psi_B}$ we have: 
\begin{equation}
    \bl Q_1^B(\lO^\dagger_A-\lO^\dagger_B)Q^{1+a}_A\br_\alpha=0,\, \alpha=1,\dots,L,\, a=1,\dots,N-1\,.
\end{equation}

Now if we insert the explicit form of $\lO^\dagger$ \eqref{dualbax} for the two states $A$ and $B$ we obtain the following system of equations:
\begin{equation}\label{linsystem}
    \displaystyle \sum_{\beta=1}^{L}\sum_{b=1}^{N-1}\bl Q_1^B u^{\beta-1} Q^{1+a\, [N-2b]}_A\br_\alpha I^{AB}_{b,\beta}=0,\, \alpha=1,\dots,L,\, a=1,\dots,N-1
\end{equation}
where we have defined $I^{AB}_{b,\beta}=(-1)^{b}(I_{b,\beta}^A - I_{b,\beta}^B)$. Here $I^A_{b,\beta}$ ($I^B_{b,\beta}$) are the eigenvalues of the integrals of motion $\hat{I}_{b,\beta}$ (defined in \eqref{integralsofmotion}) evaluated on the state $|\Psi_A\rangle$ ($|\Psi_B\rangle$). All other terms of the Baxter operator cancel out since they do not depend on the state. Since the collection of integrals of motion $I_{b,\beta}$ has non-degenerate spectrum at least one of the differences $I^{AB}_{b,\beta}$ must be non-zero for the two distinct states and so in order for the linear system \eqref{linsystem} to have a non-trivial solution we must have\footnote{A row in this matrix is labelled by the pair $(a,\alpha)$ and a column is labelled by the pair $(b,\beta)$. The pairs of indices $(a,\alpha)$ and $(b,\beta)$ are ordered lexicographically.}
\begin{equation}\label{functionalorth}
    \det_{(a,\alpha),(b,\beta)} \bl Q_1^B u^{\beta-1} Q^{1+a\, [N-2b]}_A\br_\alpha\  \propto\  \delta_{AB}\,.
\end{equation}
This is the functional orthogonality relation. It reproduces a crucial feature of the scalar product between two Bethe states, namely that it vanishes for two distinct states. In fact, it can be shown \cite{Gromov:2020fwh} to be exactly \textit{identical}  to the scalar product \eqref{SoVscalarprod} by including a state-independent normalisation $\lN$ which should be chosen to ensure that $\mathcal{M}_{0,0}=1$ and so we have
\begin{equation}\label{funscalarprod}
    \langle\Psi_A|\Psi_B\rangle = \frac{1}{\lN}\det_{(a,\alpha),(b,\beta)} \bl Q_1^B u^{\beta-1} Q^{1+a\, [N-2b]}_A\br_\alpha\,,
\end{equation}
where the normalisation factor $\lN$ is given by 
\begin{equation}
    \lN =\prod_{\alpha>\beta}(\theta_\alpha-\theta_\beta)^{N-1} = (-1)^{\frac{L}{2}(L-1)(N-1)}\Delta_\theta^{N-1}
\end{equation}
where $\Delta_\theta$ is the Vandermonde determinant
\begin{equation}
    \Delta_\theta:=\prod_{\alpha<\beta}(\theta_\alpha-\theta_\beta)\,.
\end{equation}
\subsection{Scalar product between arbitrary factorisable states}
\label{scalarprod}

The functional orthogonality relation \eqref{functionalorth}, together with the orthogonality conditions for the vacuum state $\mathcal{M}_{0,\svx}=\delta_{0,\svx}$ and $\mathcal{M}_{\svy,0}=\delta_{\svy,0}$, allows one to completely determine all matrix elements $\mathcal{M}_{\svy,\svx}$ of the measure from the knowledge of the determinant form of the scalar product \eqref{funscalarprod}. In fact, by considering all possible pairs of different states $A$ and $B$, we obtain a system of linear equations for every matrix element. A rigorous counting can even be carried out in the infinite-dimensional case \cite{Gromov:2020fwh}.

As was noticed in \cite{Gromov:2020fwh}
the fact that the determinant \eqref{funscalarprod}
reproduces the sum \eqref{SoVscalarprod} is independent of whether or not the functions $Q_1$ and $Q^{1+a}$ actually solve the Baxter equation. As a result, we can consider any so-called {\it factorisable} states $|\Phi\rangle$ and $\langle \Theta|$ with wave functions 
\begin{equation}\label{factorisable}
    \Phi(\svx)= \displaystyle \prod_{\alpha=1}^L\prod_{a=1}^{N-1} F_\alpha(\svx_{\alpha,a}),\quad  \Theta(\svy)= \displaystyle \prod_{\alpha=1}^L\displaystyle \det_{1\leq a,b\leq N-1} G_\alpha^{1+a}\left(\svy_{\alpha,b}+\frac{i}{2}(N-2)\right)\,,
\end{equation}
where $F_\alpha$ and $G_\alpha^{1+a}$ can be any functions (chosen such that the infinite sum over SoV states converges) and their scalar product will still be given by the determinant \eqref{funscalarprod}, where the bracket is understood as the sum over residues \eq{brsum}.

A useful and non-trivial example to consider is the case of the scalar product between eigenstates of two transfer matrices built with different twists. Concretely, we consider a family of transfer matrices $\T_{a}$ corresponding to \eqref{highertransfer} and another family of transfer matrices $\tilde{\T}_{a}$
with $G$ replaced by $\tilde{G}$, obtained by replacing the twist parameters $\lambda_i$ of $G$ with a new set $\tilde{\lambda}_i$. It was first demonstrated in \cite{Ryan:2018fyo} and further explored in \cite{Ryan:2020rfk,Gromov:2020fwh} that the SoV bases 
are independent of the twist parameters $\lambda_j$ after appropriate normalisation. As a result, the same SoV bases serve to factorise the wave functions of transfer matrices built with \textit{any} twist matrix of the form \eqref{companion} such as $\tilde{G}$ and so we have
\begin{equation}\label{SoVscalarprodtwist}
    \langle\Psi_A |\tilde{\Psi}_B\rangle= \displaystyle \sum_{\svx,\svy}  \Psi_A(\svy)\mathcal{M}_{\svy,\svx}\tilde{\Psi}_B(\svx)\,,
\end{equation}
where we have denoted a right eigenstate of the transfer matrices $\tilde{\T}_a$ by $|\tilde{\Psi}_B\rangle$.
This means that we can easily compute scalar products between eigenstates of transfer matrices with different twists via determinants of Q-functions. In particular we get:
\begin{equation}\la{scalar}
    \langle\Psi_A |\tilde{\Psi}_B\rangle=\frac{1}{\lN} \det_{(a,\alpha),(b,\beta)} \bl \tilde{Q}_1^B u^{\beta-1} Q^{1+a\, [N-2b]}_A\br_\alpha
\end{equation}
where $\tilde{Q}_1^B$ are the Q-functions associated to the state $|\tilde{\Psi}_B\rangle$ and the transfer matrices $\tilde{\T}_a(u)$.

\subsection{Correlators from variation of spin chain parameters}\la{sec:det}

The functional SoV approach allows one to extract a host of diagonal form-factors by varying the integrals of motion with respect to some parameter $p$ of the spin chain, such as twists $\lambda_j$ or inhomogenities $\theta_\alpha$ or even the local representation weights \cite{Cavaglia:2021mft}. The construction is based on standard quantum mechanical perturbation theory and we review it here for completeness. 

The starting point is the trivial relation $\bl Q_1 \lO^\dagger Q^{1+a}\br=0$
with $Q^{1+a}$ being on-shell Q-function i.e. satisfying the dual Baxter equation $\lO^\dagger Q^{1+a}=0$.
This obviously remains true if we consider a variation $p\rightarrow p+\delta p$ of the parameter $p$ in $Q^{1+a}$ and $\lO$ resulting in $\bl Q_1 (\lO^\dagger +\delta_p \lO^\dagger) (Q^{1+a}+\delta_p Q^{1+a})\br=0$. Expanding to first order in $\delta$, using the adjointness property
of $\lO^\dagger$ 
and also assuming that $\lO Q_1=0$ we obtain at the leading order in the perturbation
\begin{equation}
    \bl Q_1 \partial_p\lO^\dagger Q^{1+a}\br_\alpha=0\,.
\end{equation}
By expanding out $\partial_p \lO^\dagger$ this relation allows one to obtain an inhomogeneous linear system for the derivatives $\partial_p I_{b,\beta}$ of the integral of motion eigenvalues $I_{b,\beta}$. As a result we have the relation, following from Cramer's rule,
\begin{equation}\la{ratio}
    \frac{\langle\Psi|\partial_p \hat{I}_{b',\beta'}|\Psi\rangle}{\langle \Psi|\Psi\rangle} = \partial_p I_{b',\beta'} = \frac{\displaystyle \det_{(a,\alpha),(\beta,b)} m'_{(a,\alpha),(b,\beta)}}{\displaystyle \det_{(a,\alpha),(\beta,b)} m_{(a,\alpha),(b,\beta)}}\,,
\end{equation}
where $m_{(a,\alpha),(b,\beta)}=\bl Q_1 u^{\beta-1}\lD^{N-2b} Q^{1+a} \br_\alpha$ and $m'$ is obtained from $m$ by replacing the column $(b',\beta')$ with 
$y_{(a, \alpha)} \equiv\bl Q_{1} \hat{Y}_{p} \circ Q^{1+a}\br_{\alpha}$,
where $\hat{Y}_p$ is the part of $\partial_p\lO^\dagger$ which does not depend on the integrals of motion, given by:
\beq
\hat{Y}_{p}=-\left(\partial_{p} Q_{\theta}^{[-2 \bs]} \mathcal{D}^{-N}+(-1)^{N} \partial_{p} Q_{\theta}^{[+2 \bs]} \mathcal{D}^{+N}\right)-\sum_{b=1}^{N-1}(-1)^{b+1} \partial_{p} \chi_{b} u^{L} \mathcal{D}^{-2 b+N}\,.
\eeq
We introduce the short-hand notation for the determinants as follows
\beq
[o_{b,\beta}] \equiv
\det_{(a,\alpha),(b,\beta)}\bl \tilde Q_{1}^B o_{b,\beta}Q^A_{1,1+a}\br\;,
\eeq
where $o_{b,\beta}$ is some finite difference operator. Since the l.h.s. makes no reference to the twists or indices $A$ and $B$ used on the Q-functions these should be inferred from context. As such the scalar product in this notation is given by 
\begin{equation}
    \langle \Psi_A | \tilde{\Psi}_B\rangle = \frac{1}{\lN} [w^{\beta-1} \lD^{3-2b}]\,.
\end{equation}
We will also use the replacement notation 
\beqa
[(b',\beta')\to  o]\,,
\eeqa
which corresponds to replacing $w^{\beta'-1}\lD^{3-2b'}$ in the determinant $[w^{\beta-1} \lD^{3-2b}]$ with the finite difference operator $o$. For instance the numerator of \eq{ratio} becomes
\beq
\displaystyle \det_{(a,\alpha),(b,\beta)} m'_{(a,\alpha),(b,\beta)} \equiv [(b',\beta')\to  \hat Y]\;.
\eeq
Since the scalar product $\langle \Psi|\Psi\rangle$ in our normalisation is proportional to the denominator of the right hand side (see \eq{SoVscalarprodtwist})  we have 
\begin{equation}\label{diagonalff}
    \langle\Psi|\partial_p \hat{I}_{b',\beta'}|\Psi\rangle =\frac{1}{\lN}[(b',\beta')\to  \hat Y]\,.
\end{equation}
It is appealing to assume then that the operator $\d_p \hat I_{b',\beta'}$ can be characterised by this particular modification of the structure of the determinant as compared to the identity operator given by \eq{SoVscalarprodtwist}.
One can also notice that for the identity operator in 
\eq{SoVscalarprodtwist} we managed to obtain a more general relation with the left and right states corresponding to two different eigenvalues of the transfer matrix or, even more generally, to the transfer matrices with different twists. 
It is thus very tempting to upgrade the relation \eq{diagonalff} by replacing $\langle \Psi|$ and $Q^{1+a}$ accordingly by those corresponding to a different state. Whereas this does give the right result in some cases, as was noticed in \cite{Cavaglia:2018lxi}, in general this strategy, unfortunately, fails as we verified explicitly for some small length cases. 
However, for the case when the parameters $p$ are the twist angles this naive approach gives the correct result as we prove in the next section where we also provide generalisations of this result.

At this point it is, however, very easy to announce our main observation of the next section, which we prove  rigorously there. Namely, we noticed that for the case of $p=\lambda_a$
the equation \eq{diagonalff} survives a series of upgrades. Firstly, it works for two arbitrary left and right factorisable states. Secondly, and probably the most surprising, it still works for multiple derivatives in the twist parameters:
\begin{equation}\label{diagonalff1}
    \langle\Psi^A|\partial_{\lambda_{a_1}}\dots \partial_{\lambda_{a_k}} \hat{I}_{b',\beta'}|\Psi^B\rangle =\frac{1}{\lN}\[(b',\beta')\to  -\sum_{b=1}^{N-1}(-1)^{b+1} \partial_{\lambda_{a_1}}\dots \partial_{\lambda_{a_k}} \chi_{b} u^{L} \mathcal{D}^{-2 b+N}\]\,.
\end{equation}
In the next section we will derive this identity using the character projection extension of the FSoV method.
We will also see more explicitly what the operators of the type \eq{diagonalff1} are closely related with the principal operators introduced earlier.

\section{Character projection}\label{sec:sl3disc}
In this section we extend the FSoV method, introduced in the previous section, in order to obtain form-factors of non-trivial operators between two arbitrary factorisable states. 
We will use these results in the next section 
to extract the matrix elements of a set of observables in the SoV bases in a similar way to the measure, which then allows us to efficiently compute the expectation values of a complete set of physical observables.
For simplicity in this section we constrain ourselves to the $\sl(3)$ case.

\subsection{Derivation}\la{sec:der}
We start from the conjugate Baxter operator $\lO^\dagger$. A common notation for the Q-functions of the $\sl(3)$ case is $Q^{1+a}:=Q_{1,1+a}$ and we will use this notation here.  $\lO^\dagger$ gives $0$ when applied to the $Q_{1,1+a}$ functions as they satisfy the Baxter equation \eq{bax}, which in the $\sl(3)$ case becomes:
\beq\la{dBax}
\lO^\dagger = Q_\theta^{[2\bs]}\cD^3
-
\tau_1 \cD^{1}
+
\tau_2 \cD^{-1}
-\chi_{3}Q_\theta^{[-2\bs]}\cD^{-3}\;\;,\;\;
\lO^\dagger Q_{1,1+a} = 0\;.
\eeq
This implies that for any $g$, chosen such that the integral in the scalar product is convergent, we have:
\beq\label{baxtereq2}
\bl g \lO_A^\dagger Q^A_{1,a+1}\br_\alpha = 0\;\;,\;\;\alpha=1,\dots,L\;\;,\;\;a=1,2\;.
\eeq
For definiteness we take $g=\tilde Q_1^B$, which is a Q-function 
corresponding to a state of a transfer matrix with generic twist $\tilde\lambda_a$, different from that of the state $A$, which we denote as $\lambda_a$.
The corresponding characters are denoted as $\tilde \chi_r$
and $\chi_r$.
We consider the set of $2L$ equations in~\eqref{baxtereq2}
as equations on the $2L$ integrals of motion $I_{b,\beta}^A,\;b=1,2,\;\beta=1,\dots,L$, which are the non-trivial coefficients in $\tau_2(u)$ and $\tau_1(u)$. More explicitly we have
\beq\la{eqI}
\sum_{\beta,b}(-1)^{b} \bl \tilde Q_{1}^B u^{\beta-1}\lD^{3-2b} Q^A_{1,a+1}\br_\alpha
I^A_{b,\beta}
= -\sum_{r=0}^3\chi_r 
\bl \tilde Q_{1}^B \lO_{(r)}^\dagger Q^A_{1,a+1}\br_\alpha\;,
\eeq
where we introduced the following notations for the non-dynamical terms in the dual Baxter equation \eq{dBax}:
\begin{equation}
 \lO^\dagger_{(0)} = Q_\theta^{[2\bs]}\lD^3 \;\;,\;\;
 \lO^\dagger_{(1)} = -u^L\lD \;\;,\;\;
\lO^\dagger_{(2)} = u^L\lD^{-1} \;\;,\;\;
 \lO^\dagger_{(3)} = -Q_\theta^{[-2\bs]}\lD^{-3} 
\;.
\end{equation}
The solution to \eq{eqI} can be written as a ratio of determinants.  In the notations of
section~\ref{sec:det} we have
\beq\la{eqA}
I_{b',\beta'}^A=(-1)^{b'+1}
\sum_{r=0}^3\chi_r
\frac{
[(b',\beta')\to \lO^\dagger_{(r)} ]}{
[w^{\beta-1}\lD^{3-2b}]
}\;.
\eeq
At the same time, since $I_{b',\beta'}^A$ is the eigenvalue of the operator $\hat I_{b',\beta'}$ 
on the left eigenstate $\langle\Psi^A|$ 
we have
\beq\la{eqB}
I_{b',\beta'}^A=\frac{\langle \Psi^A|\hat I_{b',\beta'}|\tilde\Psi^B\rangle}
{\langle \Psi^A|\tilde \Psi^B\rangle}
={\cal N}\frac{\langle \Psi^A|\hat I_{b',\beta'}|\tilde \Psi^B\rangle}
{[w^{\beta-1}\lD^{3-2b}]
}
\eeq
where in the last identity we used the expression for the scalar product of two factorisable states \eq{scalar}. Comparing 
\eq{eqA} and \eq{eqB} we
get
\beq\la{eqII}
\langle \Psi^A|\hat I_{b',\beta'}|\tilde \Psi^B\rangle = 
\frac{(-1)^{b'+1}}{{\cal N}}
\sum_{r=0}^3\chi_r\;
[(b',\beta')\to \lO^\dagger_{(r)} ]\;.
\eeq
The next step, which we call {\it character projection} is quite crucial. As we discussed in 
Section~\ref{principalop} the IoMs, as operators, depend non-trivially on the twist of the spin chain $\lambda_a$, but when expressed in terms of the characters this dependence is linear in $\chi_r$, see \eqref{Ichi}.
We also notice that the r.h.s. of~\eqref{eqII} has explicit linear dependence on $\chi_r$. However, notice that both sides of \eq{eqII} have an additional implicit dependence on the twists through the eigenstate $\langle \Psi^A|$ and the corresponding Q-function $Q^A_{1,1+a}$. In order to remove this dependence we use the result of section~\ref{scalarprod}, which states that
the determinants in the r.h.s. of \eq{eqII} can be written in the form
\beq\la{compA}
\frac{(-1)^{b'+1}}{{\cal N}}[(b',\beta')\to \lO^\dagger_{(r)} ]
=\sum_{\svx,\svy}\Psi^A(\svy) M^{{(r);b',\beta'}}_{\svy,\svx}
\tilde\Psi^B(\svx) 
\eeq
which is analogous to \eq{SoVscalarprodtwist}, with $M^{{(r);\beta',a'}}_{\svy,\svx}$ being independent of the states $A$ and $B$.
In section~\ref{matrixelements} we compute the coefficients $M^{{(r);\beta',a'}}_{\svx,\svy}$ explicitly.
The expression \eq{compA} is obtained by 
expanding the determinant and comparing the combinations of the Q-functions with those appearing in $\tilde\Psi^B(\svx) $ and $\Psi^A(\svy)$ as shown in \eq{wave1} and \eq{wave2}.

At the same time for the l.h.s. of \eq{eqII} we have
\beq\la{compB}
\langle \Psi^A|\hat I_{b',\beta'}|\tilde\Psi^B\rangle=
\sum_{\svx,\svy}
\langle \Psi^A|\svy\rangle\langle\svy|\hat I_{b',\beta'}|\svx\rangle\langle\svx|\tilde\Psi^B\rangle
\eeq
by using completeness of SoV bases. The operator $\hat I_{b',\beta'}$ can be decomposed into terms corresponding to different characters $\chi_r$ as $\hat I_{b',\beta'}=\sum_{r=0}^3 \chi_r \hat I_{b',\beta'}^{(r)}$, see \eq{Ichi}. By comparing 
\eq{compA} and \eq{compB}
we get
\beq\la{PPxy}
\sum_{\svx,\svy}
\langle \Psi^A|\svy\rangle
\langle\svx|\tilde\Psi^B\rangle
\[\sum_{r=0}^3\chi_r\(
\langle\svx|\hat I_{b',\beta'}^{(r)}|\svy\rangle-M^{{(r);b',\beta'}}_{\svx,\svy}
\)\]=0\,.
\eeq
Note that the expression in the square brackets does not depend on the state $A$ and only carries the information on the twist of this state in the characters $\chi_r$. For simplicity, consider an arbitrary finite dimensional case with representation of dimension $D$ per site.
Considering the expression in the square bracket as a collection of $D^L\times D^L$ numbers computed for different $\svx$ and $\svy$ we get a system of linear equations on those coefficients.
There are $D^L$ states $\langle\Psi^A|$ and $D^L$ states $|\tilde\Psi^B\rangle$ so we have as many equations as unknowns and furthermore the matrix $\langle\Psi^A|\svy\rangle\langle\svx|\tilde\Psi^B\rangle$
can be considered as an overlap matrix between two complete bases $\langle \Psi^A|\otimes |\tilde\Psi^B\rangle$ to $\langle\svx|\otimes |\svy\rangle$ in the double copy of the initial Hilbert space $H\otimes H^\dagger$, and thus is not degenerate.
In fact we have many more of the equations as $|\tilde\Psi^B\rangle$ 
contains its own set of independent continuous twist parameters. We see that as a consequence of the consistency of the linear system it should have a trivial solution and thus we should have that the square bracket is identically zero
\beq\label{roundbr}
\sum_{r=0}^3\chi_r\(
\langle\svx|\hat I_{b',\beta'}^{(r)}|\svy\rangle-M^{{(r);b',\beta'}}_{\svx,\svy}
\)=0\;.
\eeq
The above equation also stays true for the infinite dimensional case and this will be argued in Appendix \ref{dict} where the coefficients $M^{{(r);b',\beta'}}_{\svx,\svy}$ are explicitly computed.

Another way to arrive to \eq{roundbr}
from \eq{PPxy} is by multiplying the l.h.s. by $\langle \svy'|\Psi^A\rangle
\langle\tilde\Psi^B|\svx'\rangle
$ and summing over complete basis of eigenstates $\Psi^A$ and $\Psi^B$
with the completeness relation \footnote{See Appendix \ref{basisapp} for a proof of the existence of this relation for our family of infinite-dimensional representations.}
\begin{equation}\label{complrel}
    1 = \sum_{A}|\Psi_A\rangle\langle\Psi_A|
\end{equation}
As a result we have $\sum_A\langle \svy'|\Psi^A\rangle \langle \Psi^A|\svy\rangle = \delta_{yy'}$ which removes the dependence on the wave functions and leads to \eq{roundbr}.

Next, the round bracket in \eqref{roundbr} does not depend on the twists, 
and the only way the above identity stays true for arbitrary values of twists is if
\beq\la{Isov}
\langle\svx|I_{b',\beta'}^{(r)}|\svy\rangle=M^{{(r);b',\beta'}}_{\svx,\svy}\;.
\eeq
Thus we get a set of $4\times 2\times L$ observables $\hat{I}_{a,\alpha}^{(r)}$ explicitly in the SoV basis, which are precisely the coefficients of the principal operators $\pr_{a,r}(u)$
\beq\la{eqpr}
\pr_{a,r}(u)=
\sum_{\beta=1}^{L} I^{(r)}_{a,\beta}u^{\beta-1}+u^L\delta_{a,r}\;.
\eeq
In section~\ref{completeness}
we prove that this set of observables is complete and we will explicitly compute the SoV matrix elements for $\pr_{a,r}(u)$ in section~\ref{principalsovbasis}.

Finally, after obtaining the relations \eqref{Isov} for the individual operators in the SoV basis we can revert the logic and multiply \eq{Isov} by $\sum_{\svx,\svy}
\langle \Psi^A|\svy\rangle
\langle\svx|\tilde\Psi^B\rangle
$ to obtain the {\it character projected} version of the equation \eq{eqII}
\beq\la{eqIIpr}
\boxed{\langle \Psi^A|\hat I^{(r)}_{b',\beta'}|\tilde \Psi^B\rangle = 
\frac{(-1)^{b'+1}}{{\cal N}}
[(b',\beta')\to \lO^\dagger_{(r)}]}\;,
\eeq
which constitutes the main result of this section.
To summarise, we obtained a determinant form of form-factors of all operators $\hat{I}^{(r)}_{b,\beta}$ between two 
arbitrary factorisable states.
It is easy to see that \eq{eqIIpr} is equivalent to \eq{diagonalff1}.

Before closing this subsection a comment is in order. A key step in our derivation relied on the denominator in \eqref{eqB} being non-zero. This is indeed non-zero as long as $|\tilde{\Psi}_B\rangle$ is not orthogonal to $\langle\Psi_A|$ which is true as long as $|\tilde{\Psi}_B\rangle$ is a generic factorisable state or as long as the twists in $|\tilde{\Psi}_B\rangle$ are independent from those in $\langle\Psi_A|$. The expressions \eqref{eqIIpr} for the form-factors are then valid for any choice of twists or indeed any factorisable states. However, it is possible to recast the derivation in an alternate way which avoids this step completely and we present it in Appendix \ref{app:alt}: the above derivation, which may be singular in certain degenerate cases, is presented to highlight the determinant origin of our result as a consequence of Cramer's rule. Finally, the counting argument presented above relied on the representation being finite dimensional. The results remain true even when extended to the infinite-dimensional case as is discussed in Appendix~\ref{dict}.

\subsection{Form-factors for $\sl(3)$ principal operators}
\label{formfactors}
In the previous section we found the form-factors of the coefficients $\hat{I}_{a,\alpha}^{(r)}$ of the $u$-expansion of the principal operators $\pr_{a,r}(u)$. In this section we derive compact determinant expressions for the form-factors of $\pr_{a,r}(u)$ themselves as functions of the spectral parameter $u$.
We will use $w$ for the dummy spectral parameter appearing inside the determinants to avoid confusion with $u$ -- the argument of $\pr_{a,r}(u)$.

Let us start from $\pr_{1,1}(u)=T_{11}(u)$. From \eq{eqpr} we see this principal operator is a generating function for the set of operators $\hat I_{1,\alpha}^{(1)}$ with $\alpha=1,\dots,L$. From \eq{eqIIpr}
we thus have
\beqa\la{sumdet}
\langle \Psi^A|T_{11}(u)|\tilde{\Psi}^B\rangle &=& 
u^L
\langle \Psi^A|\tilde{\Psi}^B\rangle
-\frac{1}{\cN}
\sum_{\beta'=1}^{L} u^{\beta'-1}
[(1,\beta')\rightarrow w^L \lD]\;.
\eeqa
This expression appears to be a sum over determinants. Let us show that it can be compressed into a single determinant.
Let us write the determinants in the sum \eq{sumdet} more explicitly by introducing the notation 
\begin{equation}
    [o_{b,\beta}] = [o_{1,1},\dots,o_{1,L},o_{2,1},\dots,o_{2,L}]\,,
\end{equation}
obtaining 
\beqa
{\cal N}\langle \Psi^A|\pr_{1,1}(u)|\tilde{\Psi}^B\rangle&=&\\
\nn&-&{\color{red}u^0}[{\color{blue}w^L\lD},w\lD,w^2\lD,\dots,
w^{L-1}\lD,
\lD^{-1},w\lD^{-1},\dots,w^{L-1} \lD^{-1}]\\
\nn&-&{\color{red}u^1}[\lD,{\color{blue}w^L\lD},w^2\lD,\dots,
w^{L-1}\lD,
\lD^{-1},w\lD^{-1},\dots,w^{L-1} \lD^{-1}]\\
\nn&-&{\color{red}u^2}[\lD,w\lD,{\color{blue}w^L\lD},\dots,
w^{L-1}\lD,
\lD^{-1},w\lD^{-1},\dots,w^{L-1} \lD^{-1}]\\
\nn&\dots\\
\nn&+&{\color{red}u^L}[\lD,w\lD,w^2\lD,\dots,
w^{L-1}\lD,
\lD^{-1},w\lD^{-1},\dots,w^{L-1} \lD^{-1}]\,,
\eeqa
where in the last term we also wrote the overlap of the states in the determinant form \eq{scalar}. By a simple rearrangement of  the columns we get
$
(-1)^{L}[\{(w^j-u^j)\lD\}_{j=1}^L,
\{w^{j-1}\lD^{-1}\}_{j=1}^L]
$
or equivalently
$
(-1)^{L}[\{(w-u) w^{j-1}\lD\}_{j=1}^L,
\{w^{j-1}\lD^{-1}\}_{j=1}^L]
$. Hence we arrive to the following expression as a single determinant
\beqa\label{T11}
\boxed{
\langle \Psi^A|\pr_{1,1}(u)|\tilde{\Psi}^B\rangle = 
\frac{(-1)^{L}}{\cN}[\{(w-u)w^{j-1}\lD\}_{j=1}^L,
\{w^{j-1}\lD^{-1}\}_{j=1}^L]\;.
}
\eeqa
We will now introduce a very convenient shorthand notation. For ordered sets ${\bf u}_a$ and $4$ integers $L_a$, $a=0,1,2,3$ we define the following object
\beqa\la{brf}
&&\Big[
L_0;{\bf u}_0
\Big|
L_1;{\bf u}_1
\Big|
L_2;{\bf u}_2
\Big|
L_3;{\bf u}_3
\Big]_\Psi=\frac{1}{\cal N}\times\\
\nn&&
\Big[\Big\{
\frac{\Delta_{{\bf u}_0\cup w}}{
\Delta_{{\bf u}_0}
} w^{j} \lD^{3}\Big\}_{j=0}^{L_0-1},
\Big\{
\frac{\Delta_{{\bf u}_1\cup w}}{
\Delta_{{\bf u}_1}
} w^{j} \lD^{1}\Big\}_{j=0}^{L_1-1},
\Big\{
\frac{\Delta_{{\bf u}_2\cup w}}{
\Delta_{{\bf u}_2}
} w^{j} \lD^{-1}\Big\}_{j=0}^{L_2-1},
\Big\{
\frac{\Delta_{{\bf u}_3\cup w}}{
\Delta_{{\bf u}_3}
} w^{j} \lD^{-3}\Big\}_{j=0}^{L_3-1}
\Big]\,,
\eeqa
where $\Delta_{\bf v}$ for some ordered set $\bf v$ is a Vandermonde determinant
\beq
\Delta_{\bf v} = \prod_{i<j} (v_i-v_j)
\eeq
and ${\bf v}\cup w$ means that we add one element $w$ to the ordered set ${\bf v}$ at the end. 
For example equation \eq{T11} can be written as
\beqa\la{P11}
\langle \pr_{1,1}(u)\rangle = 
\Big[ 0;\Big|L;u\Big |L;\Big |0;\Big]_\Psi\;.
\eeqa
Here and below we will systematically omit 
$\Psi^A$ and $\tilde \Psi^B$.
Note that the determinant in the r.h.s. of
\eq{P11} implicitly contains the Q-functions of the corresponding states.

Using a similar strategy as above we derived the following single determinant expressions for the  principal operators between two arbitrary factorisable states

\beqa
\bea{lcllllllll}
\langle \hat 1\rangle =
&\Big[&0;
&\Big|&L;
&\Big|&L;
&\Big|&0;
&\Big]_\Psi
\\ \vspace{1mm}
\langle  \pr_{1,0}
(u)\rangle = -
&\Big[&
1;\theta-i\bs
&\Big|&L-1;
u
&\Big|& 
L;
&\Big|&
0;
&\Big]_\Psi
\\ \vspace{1mm}
\langle \pr_{1,1}
(u)\rangle = 
&\Big[&0;
&\Big|&
L;u
&\Big|& 
L;
&\Big|&0;
&\Big]_\Psi
\\ \vspace{1mm}
\langle \pr_{1,2} 
(u)\rangle = (-1)^L
&\Big[&0;
&\Big|&
L-1;u
&\Big|& 
L+1;
&\Big|&0;
&\Big]_\Psi
\\ \vspace{1mm}
\langle \pr_{1,3}
(u)\rangle = 
-&\Big[&0;
&\Big|&
L-1;u
&\Big|& 
L;
&\Big|&
1;\theta+i\bs
&\Big]_\Psi
\\ \vspace{1mm}
\langle \pr_{2,0}
(u)\rangle = (-1)^L
&\Big[&
1;\theta-i\bs
&\Big|&
L;
&\Big|& 
L-1;u
&\Big|&0;
&\Big]_\Psi
\\ \vspace{1mm}
\langle \pr_{2,1}
(u)\rangle = 
(-1)^{L-1}
&\Big[&0;
&\Big|&
L+1;
&\Big|& 
L-1;u
&\Big|&0;
&\Big]_\Psi
\\ \vspace{1mm}
\langle \pr_{2,2}
(u)\rangle = 
&\Big[&0;
&\Big|&
L;
&\Big|& 
L;u
&\Big|&0;
&\Big]_\Psi
\\ \vspace{1mm}
\langle \pr_{2,3}
(u)\rangle = (-1)^L
&\Big[&0;
&\Big|&
L;
&\Big|&
L-1;u
&\Big|&1;\theta+i\bs
&\Big]_\Psi
\eea
\eeqa

Here we have defined $\theta\pm i\bs:=\{\theta_1\pm i\bs,\dots,\theta_L\pm i\bs\}$. In the next section we will use these expressions to obtain the matrix elements in the SoV basis of the principal operators.

\subsection{Form-factors for $\sl(2)$ principal operators}
In order to compare with previous results in the literature 
we also write form-factors for the principal operators in the case of the $\sl(2)$ spin chain in a form similar to those of the previous section.

We start from the $\sl(2)$ Baxter operator $\lO^\dagger=Q_{\theta}^{[2s]}\lD^2-\tau_1+\chi_2Q_{\theta}^{[-2s]}\lD^{-2}$. For the $\sl(2)$ spin chain, we only have the fundamental transfer matrix $t_1(u)$, so we only have the principal operators $\pr_{1,r}(u),\;r=0,1,2$. The notation \eq{brf} in the $\sl(2)$ case becomes
\beqa
&&\Big[
L_0;{\bf u}_0
\Big|
L_1;{\bf u}_1
\Big|
L_2;{\bf u}_2
\Big]_\Psi=\frac{1}{\cal N}\times\\
\nn&&
\Big[\Big\{
\frac{\Delta_{{\bf u}_0\cup w}}{
\Delta_{{\bf u}_0}
} w^{j} \lD^{2}\Big\}_{j=0}^{L_0-1},
\Big\{
\frac{\Delta_{{\bf u}_1\cup w}}{
\Delta_{{\bf u}_1}
} w^{j}\Big\}_{j=0}^{L_1-1},
\Big\{
\frac{\Delta_{{\bf u}_2\cup w}}{
\Delta_{{\bf u}_2}
} w^{j} \lD^{-2}\Big\}_{j=0}^{L_2-1},
\Big]\;.
\eeqa
Following exactly the same steps as for $\sl(3)$ we find that the matrix elements for the principal operators and the identity operator are given by
\renewcommand{\arraystretch}{1.7}
\beqa
\label{sl2insert}
\bea{lllllllllll}
\langle\hat 1\rangle=
& \Big[0;&\big|&L;&\big|&0; & \Big]_\Psi \\
\langle \pr_{1,0}(u)\rangle=
+\langle T_{12}(u)\rangle
=-& \Big[1;\theta-i \bs&\big|&L-1;u&\big|&0;& \Big]_\Psi \\
\langle\pr_{1,1}(u)\rangle =
+\langle T_{11}(u)\rangle = 
& \Big[0;&\big|&L;u&\big|&0;&\Big]_\Psi \\
\langle\pr_{1,2}(u)\rangle =-\langle T_{21}(u)\rangle =(-1)^L&
\Big[0;&\big|&L-1;u&\big|&1;\theta+i \bs &\Big]_\Psi\;.
\eea
\eeqa
Here we used \eqref{sl2principal} to relate principal operators with the elements of the monodromy matrix. From these equations it is already easy to see that $T_{11}(u)=\bB(u)$ is the SoV $\bf B$-operator, which acting on the factorised wave function, replaces $Q(w)\;\to\;(u-w)Q(w)$. We will analyse the action of the remaining operators on the SoV basis in the next section.

\subsection{Principal operators in SoV basis}\label{principalsovbasis}
The goal of this section is to convert the form factors we have derived in section~\ref{formfactors} to the SoV basis. The general strategy is simple: starting from a form factor $\langle \Psi^A|\hat{O}|\tilde{\Psi}^B\rangle$, for some operator $\hat{O}$, which we assume can be expressed as 
\begin{equation}
 \langle \Psi^A|\hat{O}|\tilde{\Psi}^B\rangle=   \Big[
L_0;{\bf u}_0
\Big|
L_1;{\bf u}_1
\Big|
L_2;{\bf u}_2
\Big|
L_3;{\bf u}_3
\Big]_\Psi
\end{equation}
we insert two resolutions of the identity $\sum_{\svx}|\svx\rangle\langle\svx|=\sum_{\svy}|\svy\rangle\langle\svy|=1$:
\beq
\label{eqformfactors}
\langle \Psi^A|\hat{O}|\tilde{\Psi}^B\rangle=\sum_{\svx,\svy}\langle\Psi^A|\svy\rangle\, \langle\svx|\tilde{\Psi}^B\rangle\,\langle\svy|\hat{O}|\svx\rangle\,.
\eeq
We then use~\eqref{wave1} and~\eqref{wave2} to write the r.h.s. in terms of Q-functions. Since the l.h.s. can be written in terms of determinants of Q-functions as proven in section~\ref{formfactors}, 
we can treat~\eqref{eqformfactors} as a linear system, where the unknowns are precisely the form factors in the SoV basis. It is then immediate to read off the matrix elements $\langle\svy|\hat{O}|\svx\rangle$.

It is straightforward to deduce a general formula, which we derive in Appendix~\ref{dict}, which reads 
\begin{equation}\label{psixy}
    \Big[
L_0;{\bf u}_0
\Big|
L_1;{\bf u}_1
\Big|
L_2;{\bf u}_2
\Big|
L_3;{\bf u}_3
\Big]_\Psi = \displaystyle\sum_{\svx\svy} \tilde{\Psi}_B(\svx)\Psi_A(\svy)    \Big[
L_0;{\bf u}_0
\Big|
L_1;{\bf u}_1
\Big|
L_2;{\bf u}_2
\Big|
L_3;{\bf u}_3
\Big]_{\svx\svy} 
\end{equation}
where we have introduced the notation 
\begin{equation}\label{xybracket}
    \Big[
L_0;{\bf u}_0
\Big|
L_1;{\bf u}_1
 \Big|
L_2;{\bf u}_2
\Big|
L_3;{\bf u}_3
\Big]_{\svx\svy} 
:= \left.\frac{s_{\bf L}}{\Delta_{\theta}^2}\displaystyle \sum_{k} {\rm sign}(\sigma)\prod_{\alpha,a}\frac{r_{\alpha,n_{\alpha,a}}}{r_{\alpha,0}} \frac{\Delta_{{\bf u}_b\cup \svx_{\sigma^{-1}(b)}}}{
\Delta_{{\bf u}_b}}\right|_{\sigma_{a,\alpha} = k_{a,\alpha}-m_{\alpha,a}+a}\,.
\end{equation}
The notation used here is identical to that used for the measure \eqref{measure}, with the only difference now being the sign factor $s_{\bf L}$ is defined as, for $\sl(N)$, 
\begin{equation}
    s_{\bf L} := (-1)^{\frac{LN}{4}(L-1)(N-1)+\sum_{n=0}^N \frac{L_n}{2}(L_n-1)}
\end{equation}
and now $\sigma$ in \eqref{xybracket} is a permutation of the set 
\begin{equation}\label{sigmaset}
    \{\underbrace{0,\dots,0}_{L_0}, \underbrace{1,\dots,1}_{L_1},\underbrace{2,\dots,2}_{L_2},\underbrace{3,\dots,3}_{L_3}\}
\end{equation}
and as before $\sigma_{\alpha,a}$ denotes the number in position $a+2(\alpha-1)$. 

\paragraph{Selection rules}
One can show that the SoV charge operator \eqref{SoVcharge} imposes selection rules on the states $\langle\svy|$ and $|\svx\rangle$ for which the matrix elements $\langle\svy|\hat{O}|\svx\rangle$ can be non-zero. As we explain in Appendix \ref{dict}, the overlap can only be non-zero if there exists some permutation $\rho^\alpha$ of $\{1,2\}$ such that 
\begin{equation}
     m_{\alpha,a} =n_{\alpha,\rho^\alpha_a}-\sigma_{\alpha,\rho^{\alpha}_a}-a
\end{equation}
for some fixed $\sigma$. We now sum over all values of $(\alpha,a)$ and denote the SoV charge of the state $\langle\svy|$ ($|\svx\rangle$) by ${\bf N}_\svy$ (${\bf N}_\svx$). We obtain 
\begin{equation}
    {\bf N}_\svy-{\bf N}_\svx = 3L - \displaystyle\sum_{\alpha,a}\sigma_{\alpha,\rho^\alpha_a}\,.
\end{equation}
Since $\sigma$ is a permutation of \eqref{sigmaset} the sum $\displaystyle\sum_{\alpha,a}\sigma_{\rho^\alpha_a,\alpha}$ simply equates to $L_1+2L_2+3L_3$ and hence we see that $\langle\svy|\hat{O}|\svx\rangle$ is only non-zero if 
\begin{equation}\label{selectionrule}
    {\bf N}_\svy-{\bf N}_\svx =3L -\displaystyle\sum_{n=0}^3 n\, L_n\,.
\end{equation}
Notice that this reproduces the observation of \cite{Gromov:2020fwh} that the measure $\lM_{\svy\svx}=\langle\svy|\svx\rangle$ is only non-zero if ${\bf N}_\svx={\bf N}_{\svy}$. Indeed, for the measure we have $L_0=L_3=0$ and $L_1=L_2=L$. Plugging into \eqref{selectionrule} we immediately find ${\bf N}_\svx={\bf N}_{\svy}$.

\subsubsection{$\sl(2)$ matrix elements}

Using the general formula \eqref{psixy} we will compute the SoV matrix elements of the $\sl(2)$ principal operators in order to make contact with existing results in literature. 

Modifying the notation \eqref{xybracket} to the case of $\sl(2)$ we define 
\begin{equation}\label{sl2dict}
    \Big[
L_0;{\bf u}_0
\Big|
L_1;{\bf u}_1
\Big|
L_2;{\bf u}_2
\Big]_\Psi= \displaystyle\sum_{\svx\svy} \tilde{\Psi}_B(\svx)\Psi_A(\svy)    \Big[
L_0;{\bf u}_0
\Big|
L_1;{\bf u}_1
\Big|
L_2;{\bf u}_2
\Big]_{\svx\svy}
\end{equation}
with 
\begin{equation}
      \Big[
L_0;{\bf u}_0
\Big|
L_1;{\bf u}_1
\Big|
L_2;{\bf u}_2
\Big]_{\svx\svy}= \left.\frac{s_{\bf L}}{\Delta_{\theta}}\displaystyle {\rm sign}(\sigma)\prod_{\alpha,a}\frac{r_{\alpha,n_{\alpha}}}{r_{\alpha,0}} \prod_b \frac{\Delta_{{\bf u}_b\cup \svx_{\sigma^{-1}(b)}}}{
\Delta_{{\bf u}_b}}\right|_{\sigma_{\alpha} = n_{\alpha}-m_{\alpha}+1}
\end{equation}
$\sigma$ is a permutation of the set 
\begin{equation}
        \{\underbrace{0,\dots,0}_{L_0}, \underbrace{1,\dots,1}_{L_1},\underbrace{2,\dots,2}_{L_2}\}
\end{equation}
with $\sigma_{\alpha}$ denoting the number at position $\alpha$. Notice that unlike in the higher rank case there is no sum over $k$ as only $k_\alpha=n_\alpha$ is possible.

We will now use this general formula to derive the SoV matrix elements of the $\sl(2)$ principal operators. We will begin with the operator $\pr_{1,1}(u)=T_{11}(u)$ for which we have 
\begin{equation}
    \langle \pr_{1,1}(u)\rangle = \Big[
0;
\Big|
L;u
\Big|
0;
\Big]_{\Psi}\,.
\end{equation}
In this case $\sigma$ is simply a permutation of $\{1,\dots,1\}$ and the only possibility is that it is the identity permutation with $\sigma_{\alpha}=1$. As a result we find that the non-zero matrix elements $\langle\svy|\pr_{1,1}(u)|\svx\rangle$ are given by 
\begin{equation}
    \langle\svy|\pr_{1,1}(u)|\svx\rangle = \left.\frac{1}{\Delta_\theta}\displaystyle \prod_{\alpha=1}^L (u-\svx_{\alpha})\prod_{\alpha>\beta}(\svx_\alpha-\svx_\beta)\prod_{\alpha=1}^L \frac{r_{\alpha,n_{\alpha}}}{r_{\alpha,0}}\right|_{m_{\alpha}=n_{\alpha}}\,.
\end{equation}
We then read off that \footnote{For $\sl(2)$ we see that the measure is diagonal and so $\langle\svy|\, \propto\, \langle \svx|$. We keep the notation $\langle\svy|$ in order to be consistent with higher rank.}
\begin{equation}
    \langle\svy|\pr_{1,1}(u)|\svx\rangle = \prod_{\alpha=1}^L (u-\svx_\alpha) \langle\svy|\svx\rangle 
\end{equation}
and hence the operator $\pr_{1,1}(u)=T_{11}(u)$ is diagonalised in the basis $|\svx\rangle$. This is not surprising as $T_{11}(u)$ coincides with the Sklyanin's ${\bf B}$ operator when the twist is taken to be of the form \eqref{companion}. What is remarkable is that we \textit{derived} that this operator acts diagonally on the SoV basis directly from the FSoV construction. We will later see that this persists at higher rank.

Next we examine $\pr_{1,0}(u)=T_{12}(u)$ and have
\begin{equation}
    \langle\pr_{1,0}(u)\rangle = -\Big[1;\theta-i \bs\big|L-1;u\big|0; \Big]_\Psi.
\end{equation}
Using the relation \eqref{sl2dict} we obtain 
\begin{equation}
 \Big[1;\theta-i \bs\big|L-1;u\big|0; \Big]_{\svx\svy}= \left.\frac{s_{\bf L}}{\Delta_{\theta}}\displaystyle  {\rm sign}(\sigma)  \frac{\Delta_{\theta-i\bs\cup \svx_{\sigma^{-1}(0)}}}{
\Delta_{\theta-i\bs}}\Delta_{u\cup \svx_{\sigma^{-1}(1)}}\prod_{\alpha}\frac{r_{\alpha,n_{\alpha}}}{r_{\alpha,0}}\right|_{\sigma_{\alpha} = n_{\alpha}-m_{\alpha}+1}
\end{equation}
where now $\sigma$ is a permutation of the set 
\begin{equation}
        \{0,1,\dots,1\}\,.
\end{equation}
We can characterise each $\sigma$ by the property $\sigma_\gamma=0$ for some $\gamma=1,\dots,L$ and there are $L$ such permutations. Hence, we obtain
\begin{equation}
  \langle\svy|\pr_{1,0}(u)|\svx\rangle= \left. \displaystyle\displaystyle \frac{Q_\theta^{[2\bs]}(\svx_\gamma)}{\Delta_\theta}\prod_{\alpha\neq \gamma}\frac{u-\svx_\alpha}{\svx_\gamma-\svx_\alpha}\prod_{\alpha>\beta}(\svx_\alpha-\svx_\beta)\prod_{\alpha}\frac{r_{\alpha,n_{\alpha}}}{r_{\alpha,0}}\right|_{m_\gamma=n_\gamma-1, m_\alpha=n_\alpha}
\end{equation}
where we have used that $|\sigma|=\gamma-1$. The situation with $\pr_{1,2}(u)=-T_{21}(u)$ is identical. We have
\begin{equation}
    \langle\svy|\pr_{1,2}(u)|\svx\rangle = (-1)^L
\Big[0;\big|L-1;u\big|1;\theta+i \bs \Big]_{\svx\svy}\,.
\end{equation}
Now, $\sigma_\gamma$ is a permutation of 
\begin{equation}
    \{1,\dots,1,2\}
\end{equation}
Up to the fact that now $|\sigma_\gamma|=L-\gamma$ the situation is identical to the previous case and we find 
\begin{equation}
    \langle\svy|\pr_{1,2}(u)|\svx\rangle =-\left. \displaystyle \displaystyle \frac{Q_\theta^{[-2\bs]}(\svx_\gamma)}{\Delta_\theta}\prod_{\alpha\neq \gamma}\frac{u-\svx_\alpha}{\svx_\gamma-\svx_\alpha}\prod_{\alpha>\beta}(\svx_\alpha-\svx_\beta)\prod_{\alpha}\frac{r_{\alpha,n_{\alpha}}}{r_{\alpha,0}}\right|_{m_\gamma=n_\gamma+1, m_\alpha=n_\alpha}
\end{equation}
which perfectly reproduces the well-known $\sl(2)$ results \cite{Sklyanin:1991ss}.

\subsubsection{$\sl(3)$ matrix elements - explicit example}

We now turn our attention to the matrix elements of the $\sl(3)$ principal operators. Since we have access to the general formula \eqref{psixy} we will not present the matrix elements $\langle\svy| \pr_{a,r}(u)|\svx\rangle$ for each principal operator explicitly. Instead we will demonstrate an explicit computation showing the formula \eqref{xybracket} being used in practice.

We consider an $\sl(3)$ spin chain of length $L=2$. The bases $\langle\svy|$ and $|\svx\rangle$ are labelled by non-negative integers $m_{\alpha,a}$ and $n_{\alpha,a}$ respectively, with $a,\alpha\in\{1,2\}$. Hence, we will use the notation 
\begin{equation}
    \langle\svy|:=\langle m_{1,1},m_{1,2};m_{2,1},m_{2,2}|,\quad |\svx\rangle = |n_{1,1},n_{1,2};n_{2,1},n_{2,2}\rangle\,.
\end{equation}

We will compute the following matrix element
\begin{equation}\label{overlapexample}
    \langle 3,2;0,0 |\pr_{1,0}(u) |2,1;1,0\rangle\,.
\end{equation}

The starting point is the expression 
\begin{equation}
    \langle\Psi_A|  \pr_{1,0}(u)|\tilde{\Psi}_B\rangle 
= -\Big[
1;\theta-i\bs
\Big|L-1;
u
\Big|
L;
\Big|
0;
\Big]_\Psi\,.
\end{equation}
As a result of \eqref{psixy} we see that the SoV matrix elements are given by 
\begin{equation}
    \langle\svy|\pr_{1,0}(u)|\svx\rangle = -\Big[
1;\theta-i\bs
\Big|L-1;
u
\Big|
L;
\Big|
0;
\Big]_{\svy,\svx}\,.
\end{equation}
We will use the expression obtained in \eqref{xybracket} to explicitly compute \eqref{overlapexample}. Repeating it here for convenience, \eqref{xybracket} reads
\begin{equation}
\langle \svy|\hat{O}|\svx\rangle = s_{\bf L}\left.\displaystyle \sum_{k} \frac{(-1)^{|\sigma|}}{\Delta_\theta^2}\prod_{\alpha,a}\frac{r_{\alpha,n_{\alpha,a}}}{r_{\alpha,0}} \prod_b \frac{\Delta_{{\bf u}_b\cup \svx_{\sigma^{-1}(b)}}}{
\Delta_{{\bf u}_b}}\right|_{\sigma_{\alpha,a} = k_{\alpha,a}-m_{\alpha,a}+a}\,.
\end{equation}
For the case at hand, we have $L=2$ and $L_0=L_1=1$, $L_2=2$ and $L_3=0$. Furthermore, 
\begin{equation}
    {\bf u}_0 = \theta-i\bs :=\{\theta_1-i\bs,\theta_2-i\bs\},\quad {\bf u}_1=\{u\}
\end{equation}
with both ${\bf u}_2$ and ${\bf u}_3$ empty. 

First, in order to obtain a non-zero matrix element we need to check that the SoV charges of $\langle\svy|$ and $|\svx\rangle$ satisfy the SoV charge selection rule \eqref{selectionrule} which reads
\begin{equation}\label{selection2}
     {\bf N}_\svy-{\bf N}_\svx =3L -\displaystyle\sum_{n=0}^3 n\, L_n
\end{equation}
with ${\bf N}_{\svx}=\sum_{\alpha,a}n_{\alpha,a}$ and ${\bf N}_{\svy}=\sum_{\alpha,a}m_{\alpha,a}$ and $L=2$. We have
\begin{equation}
    {\bf N}_{\svx}=2+1+1 = 4,\quad {\bf N}_{\svy} = 3+2 = 5\,.
\end{equation}
For the operator $\pr_{1,0}(u)$ we have $L_0=1$, $L_1=1$ $L_2=2$ and $L_3=3$ and hence \eqref{selection2} is satisfied. As such, $\sigma$ in \eqref{xybracket} corresponds to a permutation of 
\begin{equation}\label{sigmaset1}
    \{0,1,2,2\}\,.
\end{equation}

We now need to construct permutations of the set $\{n_{1,1},n_{1,2},n_{2,1},n_{2,2}\}$ for fixed $\alpha$. In general there are $4$ possible permutations which read 
\begin{equation}
    \begin{split}
        & \{n_{1,1},n_{1,2},n_{2,1},n_{2,2}\},\quad \{n_{1,2},n_{1,1},n_{2,1},n_{2,2}\},  \\
        & \{n_{1,1},n_{1,2},n_{2,2},n_{2,1}\},\quad \{n_{1,2},n_{1,1},n_{2,2},n_{2,1}\} 
    \end{split}
\end{equation}
but if there are degeneracies in $n_{\alpha,a}$ for fixed $\alpha$ there can be fewer permutations. In our case there are no degeneracies and we have the following permutations
\begin{equation}\label{kpermset}
\{2,1,1,0\},\quad \{1,2,1,0\},\quad \{2,1,0,1\},\quad\{1,2,0,1\}\,.
\end{equation}

The formula \eqref{xybracket} requires summing over all permutations in \eqref{kpermset} for which $\sigma_{\alpha,a} = k_{\alpha,a}-m_{\alpha,a}+a$ produces a valid permutation of \eqref{sigmaset1}. For each of the permutations in \eqref{kpermset} the corresponding $\sigma_{\alpha,a}$ are given by 
\begin{equation}
    \{0,1,2,2\},\quad \{0,1,3,3\},\quad \{-1,2,2,2\},\quad \{-1,2,1,3\}\,.
\end{equation}
Only the first set corresponds to a permutation of $\{0,1,2,2\}$, which has $|\sigma|=1$, and hence the only term in the sum over permutations of $n_{\alpha,a}$ for fixed $\alpha$ comes from $\{2,1,1,0\}$. Of course, in general there can be multiple such permutations which need to be taken into account.

From here, for this single $\sigma$, we can read off
\begin{equation}
    \svx_{\sigma^{-1}(0)}=\svx_{1,1},\quad  \svx_{\sigma^{-1}(1)}=\svx_{1,2},\quad  \svx_{\sigma^{-1}(2)}=\{\svx_{2,1},\svx_{2,2}\}
\end{equation}
which results in 
\begin{equation}
   \prod_b \frac{\Delta_{{\bf u}_b\cup \svx_{\sigma^{-1}(b)}}}{
\Delta_{{\bf u}_b}} =Q_\theta^{[2\bs]}(\svx_{1,1})(u-\svx_{1,2})(\svx_{2,1}-\svx_{2,2})\,.
\end{equation}
Finally we plug everything in, obtaining 
\begin{equation}
    \langle 3,2;0,0 |\pr_{1,0}(u) |2,1;1,0\rangle=-i(u-\theta_1-i(\bs+1))\frac{Q_\theta^{[2\bs]}(\theta_1+i(\bs+2))}{(\theta_1-\theta_2)^2}\frac{r_{1,2}}{r_{1,0}}\frac{r_{1,1}}{r_{1,0}}\frac{r_{2,1}}{r_{2,0}}\,.
\end{equation}
or more explicitly
{\small
\beqa
-\frac{8 \bs^3 (\bs+1) (2 \bs+1) \left(2 \bs-i \theta _{12}\right){}^2
   \left(1-i \theta _{12}+2 \bs\right) \left(2-i \theta _{12}+2
   \bs\right) \left(2 \bs+i \theta _{12}\right) \left(1-i \theta _{1}+\bs+i u\right)}{\theta _{12}^2 \left(i-\theta
   _{12}\right) \left(i+\theta _{12}\right){}^2 \left(\theta
   _{12}+2 i\right)}\;.
\eeqa
}
\normalsize
where we have defined $\theta_{12}=\theta_1-\theta_2$.

\section{Form-factors of Multiple Insertions}
\la{sec5} In the previous sections we derived various matrix elements of the principal operators. In this section we will extend this consideration to multiple insertions of the principal operators.

The most general case can be obtained by using the matrix elements in the SoV basis, however, this does not guarantee that the form-factor will have a simple determinant form. We consider this general case in section~\ref{matrixelements}. At the same time, for a large number of combinations of the principal operators we still managed to obtain determinant representations as we explain now.

\subsection{Antisymmetric combinations of principal operators}
The set-up in this section is similar to that of section~\ref{sec:der}. We consider the $\sl(3)$ case with two factorisable states $\langle\Psi^A|$ and $|\tilde\Psi^B\rangle$. In addition we assume that the state  $\langle\Psi^A|$ is on-shell meaning that it is an actual wave function of a spin chain and that it diagonalises the transfer matrix with twists $\lambda_a$.

Let us try to extend the previous method to general multiple insertions.
The starting point is again from \eq{eqI}, which we write below for convenience
\beq\la{eqIcopy}
\sum_{b,\beta}(-1)^{b} \bl \tilde Q_{1}^B u^{\beta-1}\lD^{3-2b} Q^A_{1,a+1}\br_\alpha
I^A_{b,\beta}
= -\sum_{r=0}^3\chi_r 
\bl \tilde Q_{1}^B \lO_{(r)}^\dagger Q^A_{1,a+1}\br_\alpha\;.
\eeq
We rewrite the above equation by modifying one term in the sum in the l.h.s. at $b,\beta=b'',\beta'''$.
Namely, we replace
$\bl \tilde Q_{1}^B 
(w^{\beta''-1}\lD^{3-2b''})
Q^A_{1,a+1}\br_\alpha$
by $\bl \tilde Q_{1}^B 
{\cal O}_{(s)}^\dagger
Q^A_{1,a+1}\br_\alpha$. In order for the equality to hold we also have to change the r.h.s. accordingly
\beqa\la{rewriteEQ}
&&\sum_{\beta,b}(-1)^{b} \bl \tilde Q_{1}^B 
\left.(w^{\beta-1}\lD^{3-2b})\right|_{w^{\beta''-1}\lD^{3-2b''}\to {\cal O}_{(s)}^\dagger}
Q^A_{1,a+1}\br_\alpha
I^A_{b,\beta}\\
\nn&&= 
-\sum_{r=0}^3\chi_r 
\bl \tilde Q_{1}^B \lO_{(r)}^\dagger Q^A_{1,a+1}\br_\alpha
+
(-1)^{b''} \bl \tilde Q_{1}^B 
\[{\cal O}_{(s)}^\dagger-(w^{\beta''-1}\lD^{3-2b''}) \]
Q^A_{1,a+1}\br_\alpha
I^A_{b'',\beta''}
\;.
\eeqa
So far this is just an innocent rewriting.
Next we treat the r.h.s. as an inhomogeneous part of the linear system on $I_{b,\beta}^A$ and apply Cramer's rule. As we have two terms in the r.h.s. of \eq{rewriteEQ} we obtain a sum of two ratios of determinants. As a result, for $b',\beta'\neq b'',\beta''$ we have
\beqa
\label{rewriteEQ2}
I^A_{b',\beta'}&=&(-1)^{b'+1}\frac{
[(b'',\beta'')\to {\cal O}_{(s)}^\dagger,(b',\beta')\to \sum_r\chi_r {\cal O}_{(r)}^\dagger]
}{[(b'',\beta'')\to {\cal O}_{(s)}^\dagger]}\\
\nn&-&(-1)^{b'+b''}
I^A_{b'',\beta''}\frac{
[(b'',\beta'')\to {\cal O}_{(s)}^\dagger,(b',\beta')\to w^{\beta''-1}{\cal D}^{3-2b''}]
}{[(b'',\beta'')\to {\cal O}_{(s)}^\dagger]}\,.
\eeqa
Notice that the term with ${\cal O}^\dagger_{(s)}$ in the r.h.s. of \eq{rewriteEQ} disappears as it produces a zero determinant in the numerator. The last term in~\eqref{rewriteEQ2} can be simplified a bit as we first replace the $(b'',\beta'')$ column with ${\cal O}^\dagger_{(s)}$
and then insert into the column
$(b',\beta')$ the exact expression which was previously at the column $(b'',\beta'')$
\beqa
I^A_{b',\beta'}&=&(-1)^{b'+1}\frac{
[(b'',\beta'')\to {\cal O}_{(s)}^\dagger,(b',\beta')\to \sum_r\chi_r {\cal O}_{(r)}^\dagger]
}{[(b'',\beta'')\to {\cal O}_{(s)}^\dagger]}\\
\nn&+&(-1)^{b'+b''}
I^A_{b'',\beta''}\frac{
[(b',\beta')\to {\cal O}_{(s)}^\dagger]
}{[(b'',\beta'')\to {\cal O}_{(s)}^\dagger]}\,.
\eeqa
Next we use the previously derived \eq{eqIIpr},
which in the new notations becomes
$\left[(b',\beta')\to \lO^\dagger_{(r)}\right] = 
{(-1)^{b'+1}}{{\cal N}}
\langle \Psi^A|\hat I^{(r)}_{b',\beta'}|\tilde \Psi^B\rangle$.
We get
\beqa
&&I^A_{b',\beta'}
\langle \Psi^A|\hat I^{(s)}_{b'',\beta''}|\tilde \Psi^B\rangle
-
I^A_{b'',\beta''}{
\langle \Psi^A|\hat I^{(s)}_{b',\beta'}|\tilde \Psi^B\rangle
}{
}\\ \nn
&&=\sum_r\chi_r \frac{(-1)^{b'+b''}}{{\cal N}}{
[(b'',\beta'')\to {\cal O}_{(s)}^\dagger,(b',\beta')\to {\cal O}_{(r)}^\dagger]
}\;.
\eeqa
Then we use that $I^A_{b',\beta'}
\langle \Psi^A|=
\langle \Psi^A|\hat I_{b',\beta'}
$ to plug the l.h.s. under one expectation value
\beqa
&&
\langle \Psi^A|\hat I_{b',\beta'}\hat I^{(s)}_{b'',\beta''}
-
\hat I_{b'',\beta''}
\hat I^{(s)}_{b',\beta'}|\tilde \Psi^B\rangle
=\sum_r\chi_r \frac{(-1)^{b'+b''}}{{\cal N}}{
[(b'',\beta'')\to {\cal O}_{(s)}^\dagger,(b',\beta')\to {\cal O}_{(r)}^\dagger]
}\;.
\eeqa
Finally, we apply the character projection trick to obtain
\beqa
&&
\boxed{
\langle \Psi^A|\hat I^{(r)}_{b',\beta'}\hat I^{(s)}_{b'',\beta''}
-
\hat I^{(r)}_{b'',\beta''}
\hat I^{(s)}_{b',\beta'}|\tilde \Psi^B\rangle
=\frac{(-1)^{b'+b''}}{{\cal N}}{
[(b'',\beta'')\to {\cal O}_{(s)}^\dagger,(b',\beta')\to {\cal O}_{(r)}^\dagger]
}}\;.
\eeqa
As before, once we have this expression we can remove the assumption that $\Psi^A$ is an on-shell and replace it by a generic factorisable state following the same argument as in section~\ref{sec3}.

Finally, the derivation we outlined above can be iterated to get the following general expression for the multiple insertions of the principal operators antisymmetrised w.r.t. the multi-indices $(b,\beta)$ 
{\small\beqa\la{multinsertion}
&&
\boxed{
\langle \Psi^A|\hat I^{(s_1)}_{[b_1,\beta_1}\dots
\hat I^{(s_k)}_{b_k,\beta_k]}
|\tilde \Psi^B\rangle
=\frac{(-1)^{b_1+\dots+b_k+k}}{k!\;{\cal N}}{
[(b_1,\beta_1)\to {\cal O}_{(s_1)}^\dagger,\dots,
(b_k,\beta_k)\to {\cal O}_{(s_k)}^\dagger
]
}}\;.
\eeqa}
Note that the r.h.s. vanishes if any of the character indices $(s_i)$ coincide. Thus in order to get a nontrivial r.h.s. we can have at most $4$ antisymmetrised principal operators for the $\sl(3)$ case and $N+1$ for general $\sl(N)$. 
The fact that the r.h.s. is antisymmetric in the character indices is also reflected on the l.h.s., where this is a consequence of the commutativity of transfer matrices. In fact, expanding the relation~\eqref{commP} in $u$ and $v$ we immediately get that $\hat{I}_{[b',\beta'}^{(r)}\hat{I}_{b'',\beta'']}^{(s)}=-\hat{I}_{[b',\beta'}^{(s)}\hat{I}_{b'',\beta'']}^{(r)}$. Since this can be done for any consecutive pair of character indices in the l.h.s. of~\eqref{multinsertion}, it follows that this quantity is completely antisymmetric in the character indices as a consequence of the RTT relations \eq{yangiangens}.

Finally, like in section~\ref{formfactors} we can convert 
the expression for the form-factor of the coefficients of the principal operators into the form-factor of the principal operators themselves. For example, we have:
\beqa\la{PPff}
{\small\bea{lcllllllll}
(-1)^L\dfrac{\langle \pr_{1,1}(u)
\pr_{1,2}(v)-
\pr_{1,1}(v)
\pr_{1,2}(u)
\rangle}{u-v} =
&\Big[&0;
&\Big|&L-1;u,v
&\Big|& 
L+1;
&\Big|&0;
&\Big]_\Psi\\
\langle \pr_{1,1}(u)
\pr_{2,2}(v)-
\pr_{2,1}(v)
\pr_{1,2}(u)
\rangle = 
&\Big[&0;
&\Big|&
L;u
&\Big|& 
L;v
&\Big|&0;
&\Big]_\Psi\\
-\langle \pr_{1,0}(u)
\pr_{2,2}(v)-
\pr_{2,0}(v)
\pr_{1,2}(u)
\rangle = 
&\Big[&
1;\theta-i\bs
&\Big|&
L-1;u
&\Big|& 
L;v
&\Big|&0;
&\Big]_\Psi\\
(-1)^{L-1}\langle \pr_{1,0}(u)
\pr_{2,3}(v)-
\pr_{2,0}(v)
\pr_{1,3}(u)
\rangle = 
&\Big[&
1;\theta-i\bs
&\Big|&
L-1;u
&\Big|& 
L-1;v
&\Big|&
1;\theta+i\bs
&\Big]_\Psi\,.
\eea}
\eeqa
For a more complicated but nice looking example of a triple insertion we get:
\beqa
\bea{lcccllllll}
\frac{\epsilon^{ijk}
\langle 
\pr_{1,1}(u_{i})
\pr_{1,2}(u_{j})\pr_{1,3}(u_{k})
\rangle}{(u_1-u_2)(u_1-u_3)(u_2-u_3)} =(-1)^L
&\Big[&0;
&\Big|&
L-2;u_1,u_2,u_3
&\Big|& 
L+1;
&\Big|&
1;\theta+i \bs
&\Big]_{\Psi}\,.
\eea
\eeqa
Notice that the second form-factor in \eq{PPff}
contains exactly the same combination that we found for the expressions for $\bB$ and $\bC$ operators in \eq{BCinP}! We will discuss the implications of this observation in section \ref{secBC}.

\subsection{Via Matrix elements in SoV basis}
\label{matrixelements}
In the above subsection we demonstrated how it is possible to write a large family of correlation functions with anti-symmetrised insertions of principal operators. However, this does not exhaust all possible correlators. On the other hand, we can in principal reduce the computation of correlators with any number of insertions to sums over products of form-factors with a single insertion by inserting a resolution of the identity over transfer matrix eigenstates. In practice this is not very useful as one would need to know the Q-functions for every state and not just those appearing in the wave functions. 

This issue can be resolved by using the matrix elements of the principal operators in the SoV bases instead. Consider the double insertion 
\begin{equation}
    \langle\Psi_A|\pr_{a,r}(u)\pr_{b,s}(v)|\tilde{\Psi}_B\rangle\,.
\end{equation}
We now consider three resolutions of the identity 
\begin{equation}
    1 = \sum_{\svx} |\svx\rangle\langle\svx| = \sum_{\svy} |\svy\rangle\langle\svy| =\sum_{\svx,\svy} |\svx\rangle\langle\svy|(\lM^{-1})_{\svy,\svx}
\end{equation}
where $(\lM^{-1})_{\svy,\svx}$ denotes the components of the inverse SoV measure $\lM$ \eqref{measure} which appears in the resolution of the identity 
\begin{equation}
    1=\sum_{\svx,\svy} |\svy\rangle\langle\svx|\lM_{\svy,\svx}\,.
\end{equation}
We insert the three resolutions into the above correlator, obtaining
\beq
    \langle\Psi_A| \pr_{a,r}(u) \pr_{b,s}(v)|\Psi_B\rangle = \displaystyle \sum_{\svx,\svx',\svy,\svy'}\Psi_A(\svy)\ \langle \svy|\pr_{a,r}(u)|\svx'\rangle\ \langle\svy'| \pr_{b,s}(v) |\svx\rangle\ (\lM^{-1})_{\svy',\svx'} \Psi_B(\svx)\,.
\eeq
At this point we see that the computation of multi-insertions becomes quite complicated. Indeed, for the rank $1$ $\sl(2)$ case the measure $\lM_{\svy,\svx}$ is diagonal and so the computation of the inverse measure $(\lM^{-1})_{\svy',\svx'}$ is trivial. For higher rank the measure is no longer diagonal and $(\lM^{-1})_{\svy',\svx'}$ needs to be computed. Nevertheless, it can be computed since $\lM_{\svy,\svx}$ is explicitly known \eqref{measure} and furthermore $\lM_{\svy,\svx}$, in an appropriate order of $\svx$ and $\svy$, is an upper-triangular block diagonal matrix  where each block is finite-dimensional even in the case of non-compact $\sl(N)$ \cite{Gromov:2020fwh}. 

\subsection{SoV ${\bf B}$ and ${\bf C}$ operators}\la{secBC}

In this section we will demonstrate that our results allow one to derive that the SoV ${\bf B}$ and ${\bf C}$ operators \eqref{BandC} are diagonalised in the SoV bases $|\svx\rangle$ and $\langle\svy|$ respectively. Structurally, the ${\bf B}$ and ${\bf C}$ operators are very similar. We recall the expressions \eqref{BCinP} which read
\begin{equation}
    \begin{split}
       &  {\bf B}(u) =\pr_{1,1}(u)\pr_{2,2}(u)-\pr_{2,1}(u)\pr_{1,2}(u) \\
       &  {\bf C}(u) =\pr_{1,1}(u)\pr_{2,2}(u+i)-\pr_{2,1}(u+i)\pr_{1,2}(u)\,.
    \end{split}
\end{equation}
Both of these expressions are special cases of the general double insertion $\pr_{1,1}(u)\pr_{2,2}(v)-\pr_{2,1}(v)\pr_{1,2}(u)$ appearing in \eqref{PPff}. We will denote this operator as $B(u,v)$, that is 
\begin{equation}
    B(u,v) = \pr_{1,1}(u)\pr_{2,2}(v)-\pr_{2,1}(v)\pr_{1,2}(u)\,.
\end{equation}
By using the relation \eqref{psixy} we can convert its matrix elements in the $\Psi$ basis in \eqref{PPff} to matrix elements in the $\svx,\svy$ basis. The result simply reads 
\begin{equation}\label{Buvxy}
     \langle \svy|B(u,v)|\svx\rangle = \left.\frac{s_{\bf L}}{\Delta_{\theta}^2}\displaystyle \sum_{k} {\rm sign}(\sigma)\prod_{\alpha,a}\frac{r_{\alpha,n_{\alpha,a}}}{r_{\alpha,0}} \Delta_{u\cup \svx_{\sigma^{-1}(1)}}\Delta_{v\cup \svx_{\sigma^{-1}(2)}}\right|_{\sigma_{a,\alpha} = k_{a,\alpha}-m_{\alpha,a}+a}
\end{equation}
where $\sigma$ is a permutation of 
\begin{equation}
    \{\underbrace{1,\dots,1}_{L},\underbrace{2,\dots,2}_{L} \}\,.
\end{equation}

We now examine the special cases $v=u$ and $v=u+i$, relevant for ${\bf B}$ and ${\bf C}$ respectively. 

\paragraph{B operator}

The crucial point is that in \eqref{Buvxy} we have that $\Delta_{u\cup \svx_{\sigma^{-1}(1)}}\Delta_{v\cup \svx_{\sigma^{-1}(2)}} = (u-\svx_{\sigma^{-1}(1)})(v-\svx_{\sigma^{-1}(2)})\Delta_1\Delta_2$ and hence, we see that, for $v=u$, we have
\beq
(u-\svx_{\sigma^{-1}(1)})(u-\svx_{\sigma^{-1}(2)})=\prod_{\alpha,a}(u-\svx_{\alpha,a})
\eeq
which is independent of $\sigma$. Hence, this factor can be pulled outside the sum over permutations and we obtain
 \begin{equation}
 \begin{split}
      \langle \svy|B(u,u)|\svx\rangle & =  \prod_{\alpha,a}(u-\svx_{\alpha,a})\left.\frac{s_{\bf L}}{\Delta_{\theta}^2}\displaystyle \sum_{k} {\rm sign}(\sigma)\prod_{\alpha,a}\frac{r_{\alpha,n_{\alpha,a}}}{r_{\alpha,0}} \Delta_1\Delta_2\right|_{\sigma_{a,\alpha} = k_{a,\alpha}-m_{\alpha,a}+a} \\ & = \prod_{\alpha,a}(u-\svx_{\alpha,a}) \langle\svy|\svx\rangle\,. 
\end{split}
\end{equation}
Hence the operator ${\bf B}(u):=B(u,u)$ acts diagonally on $|\svx\rangle$ with eigenvalue $\prod_{\alpha,a}(u-\svx_{\alpha,a})$. This coincides precisely with the spectrum of Sklyanin's ${\bf B}(u)$ operator \cite{Gromov:2020fwh}.

\paragraph{C operator}
We will now show that ${\bf C}$ is diagonalised in the $|\svy\rangle$ basis in the same manner as we did for ${\bf B}$. 
We start again from the expression:
\begin{equation}
     \langle \svy|B(u,u+i)|\svx\rangle = \left.\frac{s_{\bf L}}{\Delta_{\theta}^2}\displaystyle \sum_{k} {\rm sign}(\sigma)\prod_{\alpha,a}\frac{r_{\alpha,n_{\alpha,a}}}{r_{\alpha,0}} \Delta_{u\cup \svx_{\sigma^{-1}(1)}}\Delta_{u+i\cup \svx_{\sigma^{-1}(2)}}\right|_{\sigma_{a,\alpha} = k_{a,\alpha}-m_{\alpha,a}+a}\,.
\end{equation}
We will now show that $\Delta_{u\cup \svx_{\sigma^{-1}(1)}}\Delta_{u+i\cup \svx_{\sigma^{-1}(2)}}=\prod_{\alpha,a}(u-\svy_{\alpha,a})\Delta_1\Delta_2$. We have
\begin{equation}
    \Delta_{u\cup \svx_{\sigma^{-1}(1)}}\Delta_{u+i\cup \svx_{\sigma^{-1}(2)}} = (u-\svx_{\sigma^{-1}(1)})(u+i-\svx_{\sigma^{-1}(2)})\Delta_1\Delta_2
\end{equation}
We now examine the factor $(u-\svx_{\sigma^{-1}(1)})(u+i-\svx_{\sigma^{-1}(2)})$ which can be rewritten as 
\begin{equation}
 \prod_{\alpha,a:\sigma_{a,\alpha}=1}(u-\svx_{\alpha,a})\prod_{\alpha,a:\sigma_{a,\alpha}=2}(u+i-\svx_{\alpha,a})\,.
\end{equation}
Next, we use that $\svx_{\alpha,a}=\theta_\alpha+i(\bs+n_{\alpha,a})$ and $\svy_{\alpha,a}=\theta_\alpha+i(\bs+m_{\alpha,a}-a)$ with $n_{\alpha,a}=m_{\alpha,a}-\sigma_{a,\alpha}+a$ to obtain 
\begin{equation}
\prod_{\alpha,a:\sigma_{a,\alpha}=1}(u-\svx_{\alpha,a})\prod_{\alpha,a:\sigma_{a,\alpha}=2}(u+i-\svx_{\alpha,a})=\prod_{\alpha,a}(u-\theta_\alpha-i\bs -m_{\alpha,a}+1-a)\,.
\end{equation}
The final expression coincides with  $\prod_{\alpha,a}(u-\svy_{\alpha,a})$ which is independent of $\sigma$. Hence we obtain
\beq
\langle \svy|{\bf C}(u)|\svx\rangle:=\langle \svy|B(u,u+i)|\svx\rangle=\prod_{\alpha,a}(u-\svy_{\alpha,a})\langle\svy|\svx\rangle\,,
\eeq
meaning that the operator ${\bf C}(u)$ acts diagonally on the $\langle\svy|$ basis with eigenvalue $\prod_{\alpha,a} (u-\svy_{\alpha,a})$.

\section{Extension to $\sl(N)$ spin chains}\label{sec:slnextension}

In this section we will extend our results from the previous sections to the $\sl(N)$ case. The construction is a simple generalisation of the results in the previous sections, where we focused mainly on $\sl(2)$ and $\sl(3)$ cases.
We will briefly go through the main steps of the derivations.

\subsection{Determinant representation of form-factors}

We start again from the dual Baxter operator
\beq
\lO^\dagger_A = \sum_{a=0}^N(-1)^a \tau^A_a(u) \lD^{N-2a},\quad \lO^\dagger_A Q_A^{1+a}=0\,.
\eeq
Now we consider the usual trivial identity, where $\lO^\dagger_A$ is applied to $Q^{1+a}_A$:
\beq
\bl Q_1^B\lO_A^\dagger Q^{1+a}_A\br_{\alpha}=0\;\;, \quad a=1,\dots,N-1,\,\,\, \alpha=1,\dots,L
\eeq
Now we first expand the Baxter operator and the eigenvalues of the transfer matrices $\tau_a^A$ in the spectral parameter $u$, obtaining
\beq
\sum_{b,\beta}(-1)^{b} \bl Q_{1}^B u^{\beta-1}\lD^{N-2b} Q^{1+a}_A\br_\alpha
I^A_{b,\beta}
= -\sum_{r=0}^N\chi^A_r 
\bl Q_{1}^B \lO_{(r)}^\dagger Q^{1+a}_A\br_\alpha\;,
\eeq
where we have defined
\begin{equation}
 \lO^\dagger_{(0)} = Q_\theta^{[2\bs]}\lD^N \;\;,\;\;
 \lO^\dagger_{(r)} = (-1)^{r} u^L\lD^{N-2r},\,\,r=1,\dots,N-1 \;\;,\;\;
 \lO^\dagger_{(N)} =(-1)^{N} Q_\theta^{[-2\bs]}\lD^{-N} 
\;.
\end{equation}
Using Cramers' rule, we can compute the matrix elements of the integrals of motion exactly as in the $\sl(3)$ case leading to
\begin{equation}
    I_{b',\beta'} =(-1)^{b'+1}\frac{[(b',\beta')\rightarrow \sum_{r=0}^N \chi_r\, \lO^\dagger_{(r)}]}{[w^{\beta-1}\lD^{N-2b}]}\,. 
\end{equation}
Since $\langle \Psi_A|$ is an eigenvector of $\hat{I}_{b,\beta}$ with eigenvalue $I_{b,\beta}$ we can rewrite the above as 
\begin{equation}
    \langle\Psi_A|\hat{I}_{b',\beta'}|\tilde{\Psi}_B\rangle = \frac{(-1)^{b'+1}}{\lN}\frac{[(b',\beta')\rightarrow \sum_{r=0}^N \chi_r\, \lO^\dagger_{(r)}]}{[w^{\beta-1}\lD^{N-2b}]}\,. 
\end{equation}
The principal operator coefficients $\hat{I}^{(r)}_{b,\beta}$ are then introduced via the expansion into characters of the integrals of motion $\hat{I}_{b,\beta}$
\begin{equation}
    \hat{I}_{b',\beta'} = \displaystyle\sum_{r=0}^N \chi_r \hat{I}_{b',\beta'}^{(r)}\,.
\end{equation}
Performing character projection we then obtain the form-factors
\begin{equation}
    \langle \Psi_A|\hat{I}^{(r)}_{b',\beta'}|\tilde{\Psi}_B\rangle = 
   \frac{(-1)^{b'+1}}{\lN}[(b',\beta')\rightarrow  \lO^\dagger_{(r)}]\,.
\end{equation}
We see that this relation is identical to that of the $\sl(3)$ case \eq{eqIIpr}.

In the same way as in $\sl(3)$ we can assemble the operators $\hat{I}_{b,\beta}^{(r)}$ into the generating functions $\pr_{b,r}(u)$ -- the principal operators. The form-factor of the generating function $\pr_{b',r}(u)$ defined by \eqref{genfunc} is then given by
\begin{equation}
    \langle \Psi_A|\pr_{b',r}(u)|\tilde{\Psi}_B\rangle=\delta_{b'r}u^L [w^{\beta-1}\lD^{N-2b}] + \displaystyle \sum_{\beta'=1}^L(-1)^{b'+1} u^{\beta'-1}[(b',\beta')\rightarrow \lO^\dagger_{(r)}]\,.
\end{equation}
This result can be easily recast in determinant form using the same arguments as the $\sl(3)$ case. We introduce the notation:
\beqa\la{brfslN}
&&\Big[
L_0;{\bf u}_0
\Big|\dots
\Big|
L_N;{\bf u}_N
\Big]_\Psi=\frac{1}{\cal N}\times\\
\nn&&
\Big[\Big\{
\frac{\Delta_{{\bf u}_0\cup w}}{
\Delta_{{\bf u}_0}
} w^{j} \lD^{N}\Big\}_{j=0}^{L_0-1},
\dots,
\Big\{
\frac{\Delta_{{\bf u}_N\cup w}}{
\Delta_{{\bf u}_N}
} w^{j} \lD^{-N}\Big\}_{j=0}^{L_N-1}
\Big]\;.
\eeqa

We will write explicit expression for the form factors of type $\langle \pr_{b',r}(u)\rangle$. We have that:

\beqa
\bea{|l|lllllllllll|}
\hline
r = b' 
& &\Big[0;&
&\Big|&
\dots
&\Big|& 
(L)^r;u
&\Big|&\dots
&\Big|0;&\Big]_{\Psi} \\
\hline
r=0 &(-1)^{b'L+b'+L}&\Big[1;\theta-i \bs&
&\Big|&
\dots
&\Big|& 
(L-1)^{b'};u
&\Big|&\dots
&\Big|0;&\Big]_{\Psi}\\
\hline
r=N&(-1)^{b'+L(N-b')+N+1}
&\Big[0;&
&\Big|&
\dots
&\Big|& 
(L-1)^{b'};u
&\Big|&\dots
&\Big|1;\theta+i \bs;&\Big]_{\Psi}\\
\hline
\eea
\eeqa
\beqa
\bea{|l|llllllllllllll|}
\hline
r>b'&(-1)^{b'+r+1+L(r-b')}
&\Big[0;&\Big|& \dots &\Big|&
(L-1)^{b'};u
&
&\Big|&
\dots
&\Big|& 
(L+1)^{r};
&\Big|&\dots
&\Big|0;\Big]_{\Psi}\\
\hline
r<b' & (-1)^{b'+r+L(b'-r)}
&\Big[0;&\Big|& \dots &\Big|&
(L+1)^{r};
&
&\Big|&
\dots
&\Big|& 
(L-1)^{b'};u
&\Big|&\dots
&\Big|0;\Big]_{\Psi}\\ \hline
\eea
\eeqa
\paragraph{Multiple insertions}

The expression \eqref{multinsertion} for multiple insertions generalise without modification from the $\sl(3)$ case and we have
\begin{equation}
    \langle \Psi^A|\hat I^{(s_1)}_{[b_1,\beta_1}\dots
\hat I^{(s_k)}_{b_k,\beta_k]}
|\tilde \Psi^B\rangle
=\frac{(-1)^{b_1+\dots+b_k+k}}{k!\;{\cal N}}{
[(\beta_1,b_1)\to {\cal O}_{(s_1)}^\dagger,\dots,
(\beta_k,b_k)\to {\cal O}_{(s_k)}^\dagger
]
}
\end{equation}
As mentioned in the $\sl(3)$, the l.h.s. is anti-symmetric in character indices and so in order to get a non-zero correlator we require that $k\leq N+1$. 

\subsection{Matrix elements in SoV bases}

We can repeat the arguments from the $\sl(3)$ section to compute all form-factors of the form $\langle\svy|\pr_{a,r}(u)|\svx\rangle$. We introduce the notation 
\begin{equation}
\Big[
L_0;{\bf u}_0
\Big|\dots
\Big|
L_N;{\bf u}_N\Big]_{\svy,\svx}
\end{equation}
defined by the property 
\begin{equation}\label{xysln}
    \Big[
L_0;{\bf u}_0
\Big|\dots
\Big|
L_N;{\bf u}_N\Big]_{\Psi} = \displaystyle \sum_{\svx,\svy}\Psi_A(\svy)\Big[
L_0;{\bf u}_0
\Big|\dots
\Big|
L_N;{\bf u}_N\Big]_{\svy,\svx}\tilde{\Psi}_B(\svx)
\end{equation}
where we remind the reader that the SoV wave functions are given by 
\begin{equation}
    \Psi_A(\svy)=\displaystyle\prod_{\alpha=1}^L\det_{1\leq a,a'\leq N-1}Q_A^{1+a}(\svy_{\alpha,a'}+\tfrac{i}{2}),\quad \tilde{\Psi}_B(\svx)=\displaystyle\prod_{\alpha=1}^L \prod_{a=1}^{N-1}\tilde{Q}_1^B(\svx_{\alpha,a})\,.
\end{equation}
The explicit expression for \eqref{xysln} is worked out to be 
\begin{equation}\label{slnxyexplicit}
    \left.\frac{s_{\bf L}}{\Delta_{\theta}^{N-1}}\displaystyle \sum_{k} (-1)^{|\sigma|}\prod_{\alpha,a}\frac{r_{\alpha,n_{\alpha,a}}}{r_{\alpha,0}} \prod_b \frac{\Delta_{{\bf u}_b\cup \svx_{\sigma^{-1}(b)}}}{
\Delta_{{\bf u}_b}}\right|_{\sigma_{a,\alpha} = k_{a,\alpha}-m_{\alpha,a}+a}\,.
\end{equation}
The index $b$ takes values in the set $\{0,1,\dots,N\}$, $a\in\{1,\dots,N-1\}$ and $\alpha\in\{1,\dots,L\}$ and the summation is over all permutations $k$ of the set $\{n_{\alpha,a}\}$ for fixed $\alpha$
for which $\sigma$ defined by $\sigma_{\alpha,a}=k_{\alpha,a}-m_{\alpha,a}+a$ defines a permutation of the set 
\begin{equation}
    \{\underbrace{0,\dots,0}_{L_0},\dots, \underbrace{N,\dots,N}_{L_N} \}\,.
\end{equation}
The matrix element \eqref{slnxyexplicit} is only non-zero if the SoV charges ${\bf N}_\svx$ and ${\bf N}_{\svy}$ satisfy the relation 
\begin{equation}
    {\bf N}_{\svy}-{\bf N}_{\svx} = \frac{N}{2}(N-1)L - \displaystyle \sum_{n=0}^N n\, L_n\,.
\end{equation}
The details of the derivation are exactly the same as in the $\sl(3)$ case described in Appendix~\ref{dict}. 

\paragraph{B and C operators.}

Having access to the complete set of SoV matrix elements it is now easy to determine which operators correspond to the SoV ${\bf B}$ and ${\bf C}$ operators. Following the derivation in the $\sl(3)$ case it is trivial to work out that ${\bB}(u)$ corresponds to the operator with 
\begin{equation}\label{slnB}
    {\bf u}_0 = {\bf u}_N=\{\},\quad {\bf u}_r =\{u\},\quad r=1,\dots,N-1
\end{equation}
whereas ${\bf C}(u)$ corresponds to the operator with 
\begin{equation}\label{slnC}
    {\bf u}_0 = {\bf u}_N=\{\},\quad {\bf u}_r =\{u+i(r-1)\},\quad r=1,\dots,N-1\,.
\end{equation}
Indeed, by examining the matrix element \eqref{slnxyexplicit} as in the $\sl(3)$ case we immediately read off that the operator defined by \eqref{slnB} (\eqref{slnC}) acts diagonally on $|\svx\rangle$ ($\langle\svy|$) with eigenvalue given by $\prod_{\alpha,a}(u-\svx_{\alpha,a})$ ($\prod_{\alpha,a}(u-\svy_{\alpha,a})$) and hence coincides with ${\bf B}(u)$ (${\bf C}(u)$) respectively due to the non-degeneracy of these operators' spectra. It is possible to work out what these operators correspond to in terms of principal operators $\pr_{a,r}(u)$. They are given by 
\begin{equation}
\begin{split}
    & {\bf B}(u)=(N-1)!\varepsilon^{a_1\dots a_{N-1}}\pr_{a_1,1}(u)\dots \pr_{a_{N-1},N-1}(u) \\
    & {\bf C}(u)=(N-1)!\varepsilon^{a_1\dots a_{N-1}}\pr_{a_1,1}(u)\dots \pr_{a_{N-1},N-1}(u+i(N-2))\,.
\end{split}
\end{equation}
The fact that these operators coincide with the ${\bf B}$ and ${\bf C}$ operators of \cite{Gromov:2020fwh} is not manifest -- application of the RTT relation \eqref{yangiangens} is required as was already demonstrated in the $\sl(3)$ case. Nevertheless, the fact that their spectra and eigenstates coincide guarantees that they are equal.

\section{Properties of principal operators}
\la{sec7} The main goal of this section is to demonstrate the completeness of the set of the principal operators. We show that any element of the Yangian can be obtained as a combination of the principal operators, which in at least finite dimensional cases guarantees that 
all physical observable can be obtained in this way. In the last section we also give explicit expressions for the principal operators in the diagonal frame -- i.e. in the case when the twist matrix becomes diagonal.

\subsection{Completeness}
\label{completeness}
In this section we will demonstrate a crucial property of the operator basis, namely that knowledge of the matrix elements of each of our principal operators is equivalent to the knowledge of the matrix elements of every operator $T_{ij}(u)$ in the Yangian algebra \eqref{yangiangens}. More precisely we will show that any element of the Yangian $T_{ij}(u)$
can be constructed as a polynomial of degree at most $N+1$ in principal operators.

Knowing all $T_{ij}(u)$ is essentially equivalent to the full algebra of observables.
For example, in the finite dimensional case i.e. when $\bs = -n/2,\;n\in {\mathbb Z}_+$ one can use the ``inverse scattering transform" \cite{Maillet:1999re} to construct local symmetry generators acting on a single site of the chain in terms of $T_{ij}(u)$.
The precise notion of completeness could be ambiguous --  in order to be precise in this paper when referring to completeness of the system of principal operators we understand that any element of the Yangian can be generated in  finitely many steps (independently of the length of the chain). Note that while we are not aware of any simple way to extract local operators in the infinite-dimensional case in terms of $T_{ij}(u)$ we would like to stress that these operators still contain all information about the system. For example, consider the infinite-dimensional highest-weight representation used in this paper and consider some local operator $\mathbb{E}^{(\alpha)}$. The key point is the existence of the SoV basis $\langle\svx|$ which is constructed by action of polynomials in $T_{ij}(u)$ on the SoV ground state $\langle 0|$ \cite{Gromov:2020fwh}. Hence, the action of $\mathbb{E}^{(\alpha)}$ on the SoV basis can be re-expressed as a sum over (finitely many\footnote{There are only finitely many states of a given SoV charge, and each local Lie algebra generator raises or lowers the SoV charge by some finite amount.}) SoV basis states $\langle\svx'|$ and hence the matrix elements of $\mathbb{E}^\alpha$ are completely fixed by the SoV matrix elements of the monodromy matrix $T_{ij}(u)$.

We now show that the principal operators generate the full Yangian. Our starting point is the large $u$ expansion of the operators $T_{ij}(u)$
\begin{equation}
    T_{ij}(u) = u^L\delta_{ij} + u^{L-1} \left(i \lE_{ji} - \delta_{ij}\Theta \right)+ \mathcal{O}\left(u^{L-2} \right), \quad \Theta:=\displaystyle \sum_{\alpha=1}^L \theta_\alpha\,.
\end{equation}
Note that the indices on $\lE$ are swapped compared to those on $T$. The operators $\lE_{ij}$ are generators of the global $\gl(N)$ algebra 
\begin{equation}\label{globalops22}
    \lE_{ij} = \displaystyle \sum_{\alpha=1}^L \ee_{ij}^{(\alpha)}
\end{equation}
and satisfy the $\gl(N)$ commutation relations 
\beq\la{comgln}
[\lE_{ij},\lE_{kl}]=\delta_{jk}\lE_{il} - \delta_{li}\lE_{kj}\;.
\eeq
We will now prove the following property: that any $T_{ij}(v)$ can be expressed as a commutator of a global $\gl(N)$ generator and a principal operator $T_{k1}(v)$. The key point is the RTT relation \eqref{yangiangens} expanded at large $u$ which reads
\begin{equation}
    [\lE_{ji},T_{kl}(v)]= T_{kj}(v)\delta_{il}-T_{il}(v)\delta_{kj}\,.
\end{equation}
From here it is clear that we can write any operator $T_{ij}(v)$ as
\begin{equation}\label{globcom}
    T_{ij}(v) = T_{11}(v)\delta_{ij} + [\mathcal{E}_{j1},T_{i1}(v)] = \pr_{1,1}(v)\delta_{ij} +(-1)^{i-1}[\mathcal{E}_{j1},\pr_{1,i}(v)]
\end{equation}
where the r.h.s. only contains principal operators and global $\gl(N)$ generators.

The family of principal operators includes the following global Lie algebra generators: $\lE_{1j}$ and $\lE^-=\displaystyle\sum_{j=1}^{n-1}\lE_{j+1,j}$. These appear in the asymptotics of the generating functions 
\begin{equation}
    \pr_{1,0}(u) = i u^{L-1} \lE^- + \mathcal{O}(u^{L-2}),\quad (-1)^j\pr_{1,j}(u) = u^L\delta_{j1} + u^{L-1} \left(i \lE_{1j} - \delta_{ij}\Theta \right)+ \mathcal{O}\left(u^{L-2} \right)\,.
\end{equation}
Hence, if we can prove that these operators can be used to generate the set of $\mathcal{E}_{j1}$ then it follows from \eqref{globcom} that knowing the matrix elements of all principal operators implies knowledge of the matrix elements of all $T_{ij}(u)$. From the commutation relations \eq{comgln} it is easy to see that
\beq
\lE_{j+1,1}=[\lE^-,\lE_{j1}]\;.
\eeq
Thus, we have 
\begin{equation}
    \lE_{j1} = \underbrace{[\lE^-,[\lE^-,[\dots,[\lE^-,\lE_{11}]]}_{j-1}\,,
\end{equation}
where the r.h.s. contains only principal operators.
After that from \eq{globcom} we get all operators $T_{ij}(u)$
generated, which completes the proof.

Let us remark that despite the abundance of literature on SoV in $\sl(2)$ spin chains the relation \eqref{globcom} does not seem to have been exploited. Indeed, the standard approach is to obtain the matrix elements of the one non-principal operator $T_{22}(u)$ in terms of the principal operators via the quantum determinant relation 
\begin{equation}
    {\rm qdet}T(u)= T_{11}^-T_{22}^+ - T_{21}^- T_{12}^+
\end{equation}
together with the known eigenvalue of the quantum determinant and the fact that $T_{11}(u)$ is invertible, see for example \cite{Niccoli:2020zla}. This produces a rather complicated expression for $T_{22}(u)$. On the other hand, using the relation \eqref{globcom} we see that $T_{22}(u)$ can be written in terms of principal operators simply as 
\begin{equation}
    T_{22}(u) = \pr_{1,1}(u) -[\mathcal{E}_{21},\pr_{2,1}(u)]\,.
\end{equation}

\subsection{Principal operators in the diagonal frame}

In the main part of the paper we used the frame with the twist matrix $G$ being of the special form \eq{companion}. Whereas for SoV approach this choice is extremely beneficial, as the SoV basis does not depend on the twist eigenvalues $\lambda_a$, it is not the most commonly used in the literature. A more standard choice is the diagonal twist $g={\rm diag}(\lambda_1,\dots,\lambda_N)$.
In this section we give an explicit way to relate those two conventions. 
As we will see the basic consequence of changing the frame is that the explicit expressions for the principal operators $\pr_{r,s}$ in terms of the monodromy matrix elements $T_{ij}$ will slightly change in the frame where the twist matrix is diagonal.

In the companion twist frame the transfer matrix $\T_1(u)$ is given by $\T_1(u) = {\rm tr}\left( T(u) G\right)$ where $G$ is the companion twist matrix \eqref{companion}. We want to perform a similarity transformation $\Pi(S)$ on the Hilbert space of the spin chain where $S$ is some $GL(N)$ group element and $\Pi(S)$ denotes its representative on the spin chain so that the transfer matrix transforms as ${\rm tr}\left( T(u) G\right)\rightarrow {\rm tr}\left( T(u) g\right)$ where $g$ is the diagonal twist matrix with the same eigenvalues as $G$. As was established in \cite{Gromov:2020fwh} a possible choice for $S$ is given by the Vandermonde matrix 
\begin{equation}
    (S^{-1})_{ij}= \lambda_j^{N-i}\,.
\end{equation}
Under this transformation the monodromy matrix elements $T_{ij}(u)$ transform as 
\begin{equation}\la{Ttrans}
    T_{ij}(u)\rightarrow \Pi^{-1}T_{ij}(u)\Pi = (S^{-1}T(u) S)_{ij}\;\;,\;\;\Pi\equiv \Pi(S)
\end{equation}
with similar expressions holding for anti-symmetric monodromy matrices. 

To summarise we have the wave-functions in the diagonal frame related to the wave-functions in the companion frame by
\beq
|\Psi^{\rm diag}\rangle = \Pi^{-1}|\Psi\rangle\;\;,\;\;
\langle \Psi^{\rm diag}| = \langle\Psi|\Pi\;.
\eeq
and they diagonalise the transfer matrices
${\mathbb T}_1^{{\rm diag}}(u)$ and ${\mathbb T}_1(u)$ correspondingly related as
\beq
{\mathbb T}_1^{{\rm diag}}(u) = 
\Pi^{-1}{\mathbb T}_1(u) \Pi= 
{\rm tr} (S^{-1} T(u) S G)= {\rm tr} ( T(u) g )\;.
\eeq
Similarly we define $\pr^{\rm diag}_{a,r}=
\Pi^{-1}\pr_{a,r} \Pi$ so that
\beq
\langle \Psi_A^{\rm diag}
|\pr^{\rm diag}_{a,r}
|\Psi_B^{\rm diag}\rangle = 
\langle \Psi_A|\pr_{a,r}
|\Psi_B\rangle = {\rm determinant}\,.
\eeq
Note that the above expression only holds for the states with the same twist unlike the expressions in the companion twist frame which hold for any twist on either state.

In general the expressions for the principal operators in the diagonal frame in terms of $T_{ij}$ are quite bulky, but straightforward to work out from \eq{Ttrans}.
For example for $\sl(3)$ we have 
$\pr^{\rm diag}_{1,1}=(S^{-1} T S)_{1,1}$
\beqa
\pr^{\rm diag}_{1,1}&=&
\frac{\lambda _1^2 T_{11}}{\left(\lambda _1-\lambda
   _2\right) \left(\lambda _1-\lambda
   _3\right)}-\frac{\lambda _1^2 T_{12}}{\left(\lambda
   _1-\lambda _2\right) \left(\lambda _2-\lambda
   _3\right)}+\frac{\lambda _1^2 T_{13}}{\left(\lambda
   _1-\lambda _3\right) \left(\lambda _2-\lambda
   _3\right)}\\ \nn
&+&\frac{\lambda _2^2 T_{21}}{\left(\lambda
   _1-\lambda _2\right) \left(\lambda _1-\lambda
   _3\right)}-\frac{\lambda _2^2 T_{22}}{\left(\lambda
   _1-\lambda _2\right) \left(\lambda _2-\lambda
   _3\right)}+\frac{\lambda _2^2 T_{23}}{\left(\lambda
   _1-\lambda _3\right) \left(\lambda _2-\lambda
   _3\right)}\\ \nn
&+&\frac{\lambda _3^2 T_{31}}{\left(\lambda
   _1-\lambda _2\right) \left(\lambda _1-\lambda
   _3\right)}-\frac{\lambda _3^2 T_{32}}{\left(\lambda
   _1-\lambda _2\right) \left(\lambda _2-\lambda
   _3\right)}+\frac{\lambda _3^2 T_{33}}{\left(\lambda
   _1-\lambda _3\right) \left(\lambda _2-\lambda
   _3\right)}\;.
\eeqa
Note that whereas in the companion twist frame the principal operators by definition where independent of the twist eigenvalues, in the diagonal frame they explicitly depend on $\lambda_i$.
In order to get nice looking expressions is it better to introduce the notation $T^{\rm good}(u)=S^{-1}T(u)SG=S^{-1}T(u)gS$ going back to~\cite{Gromov:2016itr}. It obeys ${\mathbb T}_1^{\rm diag}={\rm tr}(T^{\rm good})$ and is related in a simple way to the principal operators in the diagonal frame \eqref{companion} so that 
\begin{equation}
\begin{split}\la{PTg}
  &  \pr^{\rm diag}_{1,i}(u) =(-1)^{N-i} \frac{T^{\rm good}_{i,N}(u)}{\chi_N},\quad i=1,2,\dots,N\,, \\
  & \pr^{\rm diag}_{1,0}(u) = \sum_{i=1}^{N-1}\(
  T^{\rm good}_{i,i}-(-1)^{N-i}\frac{\chi_i}{\chi_N}T^{\rm good}_{i,N}
  \)\,.
\end{split}
\end{equation}
One can check that $\sum_{r=0}^N\pr^{\rm diag}_{1,r}(u)\chi_r = {\rm tr}(T^{\rm good})$. In particular, from \eq{PTg} the above we see that the form-factor of any $T^{\rm good}_{i,N}$ in the diagonal frame is a determinant.
For the particular case of $\sl(2)$ these operators generalise the well-known operators $T^{\rm good}_{11}$ and $T^{\rm good}_{22}$ which act as conjugate momenta of the separated variables encoded in $T^{\rm good}_{12}$, see \cite{Sklyanin:1984sb}.

\section{Outlook}

In this paper we used the functional separation of variables (FSoV) technique in combination with the novel character projection (CP) method to compute all matrix elements of the set of principal operators which in particular includes some individual monodromy matrix elements $T_{ij}$ and their combinations in a concise determinant form.
We also showed that they generate a complete basis of observables of the spin chain and contain the SoV $\bf{B}$ operator as a particular case.
Thus we gained access to the matrix elements of a set of operators which generates a complete set of observables in high-rank integrable $\sl(N)$ spin chains.

Let us note that determinant representations for form factors of some $T_{ij}$ have appeared in the literature before for the $\sl(3)$ case in the Nested Bethe Ansatz approach \cite{Belliard:2012av,pakuliak2014form,Pakuliak:2014ela}. However,
in addition to giving an alternative form for those objects, the results presented in this paper have a number of advantages and conceptual differences:
\begin{itemize}
    \item Firstly, the form factors are expressed directly in terms of Baxter Q-functions instead of Bethe roots. From a direct calculational perspective Q-functions offer a significant advantage \cite{Marboe:2016yyn}.
    \item  Secondly, the FSoV approach, which we use and extend here, does not require the existence of a highest-weight state. As such our approach is applicable to models which do not have the highest-weight state, for example the conformal spin chain (fishchain)\footnote{The fishchain captures the operators with non-trivial dimension at finite coupling, but it misses a large class of protected operators, which in turn are governed by a different integrability construction known as the (hyper)-eclectic spin chain, see \cite{Ipsen:2018fmu,Ahn:2020zly,Ahn:2021emp,Garcia:2021mzb}.} \cite{Gromov:2019bsj} describing correlators with non-trivial coupling dependence in $4$D conformal fishnet theory.
    \item Thirdly, as demonstrated, our approach is valid for any rank $\sl(N)$ with general formulas being almost equally simple to write down as for $\sl(3)$.
    \item Fourthly, our formulas are applicable to the set-up where the transfer matrix eigenstates are constructed with two distinct twists, which have attracted attention recently \cite{belliard2021overlap}, or in fact any two arbitrary off-shell states (to which we refer to as factorisable). In addition to being a new result, this is a very important technical advantage for example in non-highest-weight models where the scalar product between states built with the same twist is divergent \cite{Cavaglia:2021mft} and so deforming one set of twists serves as a natural regulator\footnote{See  \cite{Cavaglia:2020hdb} for the explicit realisation of the twist in fishnet CFT.}. 
    \item Finally, using our approach we were able to compute the matrix elements of the principal operators in the SoV bases meaning one can compute the matrix elements of any number of insertions. 
\end{itemize}

The FSoV approach was already worked out in detail in \cite{Cavaglia:2021mft} and thus the new methods we developed here can be applied immediately as the CP method is very general.
Furthermore, the operator SoV construction for any highest-weight representation was carried out in \cite{Ryan:2018fyo,Ryan:2020rfk} which can then be combined with the FSoV method to extract the SoV measure in the same way as in \cite{Gromov:2020fwh} and then the form factors as in this work.

We also believe that the formulas presented for form-factors extend immediately to the $q$-deformed high-rank XXZ case \cite{Maillet:2018rto} after simple modification as is already the case in the $\sl(2)$ setting \cite{Niccoli:2012ci}, and it would be interesting to check directly, allowing one to extend the recent rank $1$ results \cite{Pei:2020ljw} and to study high-rank correlators at zero temperature along the lines of \cite{Niccoli:2020zla}. From our results it is also possible to extract form-factors of local operators using the quantum inverse transform \cite{Maillet:1999re}. From here there are many interesting directions to pursue. For example, these can be used in the computation of current operators \cite{Pozsgay:2019xak} which have numerous applications. It would also be very interesting to compare with the results of \cite{Pozsgay:2019ekd}, where certain mean-values related to current operators are shown to factorise, and to understand such results from an SoV perspective.

Another interesting direction would be to construct the so-called ${\bf A}$ operator \cite{Sklyanin:1992sm} of Sklyanin, which is expected to act as a raising / lowering operator on the SoV bases, as some combination of principal operators as was done here for the ${\bf B}$ and ${\bf C}$ operators. This would solve an important puzzle as it is known that Sklyanin's construction is singular for the highest-weight representations we consider here, see for example the discussion in \cite{Maillet:2018bim}, and we plan to return to this question in the future.

Finally, it would be very interesting to develop the FSoV formalism and the approach to correlators developed in this paper for spin chains based on different algebras and with different boundary conditions. The Q-system for models with orthogonal symmetry has attracted huge attention recently \cite{Ferrando:2020vzk,Ekhammar:2020enr,Ekhammar:2021myw,Tsuboi:2021xzl} and will likely play a large role in the SoV approach to correlators in conformal fishnet theories in $D\neq 4$ \cite{Kazakov:2018qbr,Basso:2019xay}. As well as this the SoV construction for models with open boundary conditions has recently been studied \cite{Maillet:2019hdq} with operatorial methods. Such a spin chain is known to describe a Wilson loop in $\lN=4$ SYM with insertions of local operators in the ladders limit \cite{Gromov:2021ahm}. Another extension would be to study integrable boundary states within the SoV formalism. First steps were already taken in~\cite{Caetano:2020dyp,Cavaglia:2021mft,Gombor:2021uxz} and recently this problem received increased interest~\cite{Jiang:2020yk, Komatsu:2020yk, Gombor:2020cs,Caetano:2021dbh,Kristjansen:2021xno}. 

\paragraph{Acknowledgements}

We are grateful to A.~Cavagli\`{a} for discussions. We are also grateful to F.~Levkovich-Maslyuk for carefully reading the manuscript. The work of N.G. and P.R was supported by the European Research Council (ERC) under the European Union’s Horizon 2020 research and innovation programme
– 60 – (grant agreement No. 865075) EXACTC.

\appendix

\section{Alternative derivation}\la{app:alt}

In this appendix we present an alternative derivation of \eqref{eqIIpr} which avoids using Cramer's rule and hence avoids expressing the integral of motion eigenvalues $I_{b',\beta'}^A$  as a ratio with a potentially vanishing denominator. 

Our starting point is the following trivial equality 
\begin{equation}
    [(b',\beta')\rightarrow \lO_A^\dagger]=0\,.
\end{equation}
We then expand out $\lO^\dagger_A$ 
\begin{equation}
    \lO_A^\dagger = \displaystyle \sum_{b,\beta}(-1)^b I_{b,\beta}^A w^{\beta-1}D^{N-2b} + \sum_{r=0}^N \chi_r \lO^\dagger_{(r)} 
\end{equation}
and notice a number of cancellations. Indeed, in the sum
\begin{equation}\label{appdetsum}
    \displaystyle \sum_{b,\beta}(-1)^b I_{b,\beta}^A[(b',\beta')\rightarrow w^{\beta-1} D^{N-2b}]
\end{equation}
only a single term will survive and it is precisely $(-1)^{b'} I_{b',\beta'}^A[(b',\beta')\rightarrow w^{\beta'-1} D^{N-2b'}]$. This is a result of the anti-symmetry of the determinant as all other terms in the sum \eqref{appdetsum} will produce two identical columns in the determinant and hence vanish. 

As such we obtain the relation 
\begin{equation}
    (-1)^b [(b',\beta')\rightarrow w^{\beta'-1} D^{N-2b'}]I^A_{b',\beta'} = - \displaystyle \sum_{r=0}^{N}[(b',\beta')\rightarrow \lO^\dagger_{(r)}]
\end{equation}
and see that the coefficient of $I_{b',\beta'}^A$ is precisely $(-1)^{b'}\lN \langle\Psi_A|\tilde{\Psi}_B\rangle$. From here on the derivation proceeds exactly as in section \ref{sec:sl3disc}. Since $I^A_{b',\beta'}$ is the eigenvalue of the integral of motion $\hat{I}_{b',\beta'}$ on the state $\langle\Psi_A|$ we can replace $\langle\Psi_A|\tilde{\Psi}_B\rangle I^A_{b',\beta'}$ with $\langle\Psi_A|\hat{I}_{b',\beta'}|\tilde{\Psi}_B\rangle$. Then, we expand $\hat{I}_{b',\beta'}$ into a sum over characters $\chi_r$ to obtain
\begin{equation}
   \displaystyle \sum_{r=0}^N \chi_r \langle\Psi_A|\hat{I}_{b',\beta'}|\tilde{\Psi}_B\rangle=\frac{(-1)^{b'+1}}{\lN} \displaystyle \sum_{r=0}^{N}[(b',\beta')\rightarrow \lO^\dagger_{(r)}]\,.
\end{equation}
Finally, we equate the coefficients of the characters $\chi_r$ on both sides (character projection), which was justified in section \ref{sec:sl3disc}, and obtain the result
\begin{equation}\label{ffformapp}
    \langle\Psi_A|\hat{I}_{b',\beta'}^{(r)}|\tilde{\Psi}_B\rangle = \frac{(-1)^{b'+1}}{\lN} [(b',\beta')\rightarrow \lO^\dagger_{(r)}]\,.
\end{equation}
The derivation presented here is valid for any transfer matrix eigenstate $\langle\Psi_A|$ and any factorisable state $|\tilde{\Psi}_B\rangle$. These classes of states are enough to completely constrain the SoV matrix elements $\langle\svy|\hat{I}_{b',\beta'}^{(r)}|\svx\rangle$ (as is proven in Appendix \ref{dict}) and hence the formula \eqref{ffformapp} is valid for any two factorisable states $\langle\Psi_A|$ and $|\tilde{\Psi}_B\rangle$.

\section{Mapping $\langle\Psi_A|\hat{O}|\Psi_B\rangle$ to  $\langle\svy|\hat{O}|\svx\rangle$}\label{dict}
Our goal in this section is to prove the relation \eqref{psixy} which we repeat here for convenience 
\begin{equation}\label{genexpress}
    \Big[
L_0;{\bf u}_0
\Big|
L_1;{\bf u}_1
\Big|
L_2;{\bf u}_2
\Big|
L_3;{\bf u}_3
\Big]_\Psi = \displaystyle\sum_{\svx\svy} \tilde{\Psi}_B(\svx)\Psi_A(\svy)    \Big[
L_0;{\bf u}_0
\Big|
L_1;{\bf u}_1
\Big|
L_2;{\bf u}_2
\Big|
L_3;{\bf u}_3
\Big]_{\svx\svy} 
\end{equation}
where we use the notation 
\beqa
&&\Big[
L_0;{\bf u}_0
\Big|
L_1;{\bf u}_1
\Big|
L_2;{\bf u}_2
\Big|
L_3;{\bf u}_3
\Big]_\Psi=\frac{1}{\cal N}\times\\
\nn&&
\Big[\Big\{
\frac{\Delta_{{\bf u}_0\cup w}}{
\Delta_{{\bf u}_0}
} w^{j} D^{3}\Big\}_{j=0}^{L_0-1},
\Big\{
\frac{\Delta_{{\bf u}_1\cup w}}{
\Delta_{{\bf u}_1}
} w^{j} D^{1}\Big\}_{j=0}^{L_1-1},
\Big\{
\frac{\Delta_{{\bf u}_2\cup w}}{
\Delta_{{\bf u}_2}
} w^{j} D^{-1}\Big\}_{j=0}^{L_2-1},
\Big\{
\frac{\Delta_{{\bf u}_3\cup w}}{
\Delta_{{\bf u}_3}
} w^{j} D^{-3}\Big\}_{j=0}^{L_3-1}
\Big]\;,
\eeqa
and
\begin{equation}
    \Big[
L_0;{\bf u}_0
\Big|
L_1;{\bf u}_1
\Big|
L_2;{\bf u}_2
\Big|
L_3;{\bf u}_3
\Big]_{\svx\svy}:= \left.\frac{s_{\bf L}}{\Delta_{\theta}^2}\displaystyle \sum_{k} {\rm sign}(\sigma)\prod_{\alpha,a}\frac{r_{\alpha,n_{\alpha,a}}}{r_{\alpha,0}} \prod_b \frac{\Delta_{{\bf u}_b\cup \svx_{\sigma^{-1}(b)}}}{
\Delta_{{\bf u}_b}}\right|_{\sigma_{a,\alpha} = k_{a,\alpha}-m_{\alpha,a}+a}\,.
\end{equation}
Our starting point is the l.h.s. of~\eqref{genexpress}. By explicitly writing each entry of the matrix we can pull out the measure factors $\mu_\alpha(\omega_{\alpha,a})$ and Q-functions $\tilde{Q}^B_1$ associated to the state $|\tilde{\Psi}_B\rangle$, as the finite-difference operators in the determinant do not act on them. Hence we obtain
\begin{equation}\label{integral3}
    \displaystyle \int t(\{w_{\alpha,a}\}) \displaystyle \prod_{\alpha,a} \tilde{Q}_1^B(w_{\alpha,a})\mu_\alpha(w_{\alpha,a}){\rm d}w_{\alpha,a}
\end{equation}
where 
\begin{equation}
    t(\{w_{\alpha,a}\}) = \det_{(\alpha,a),(b,\beta)} f_b(w_{\alpha,a})w_{\alpha,a}^{\beta-1}Q_{1,1+a}\left( w_{\alpha,a}+\frac{i}{2}(3-2b)\right)\,,
\end{equation}
and 
\begin{equation}\label{fbdef}
   f_b(w)= \frac{\Delta_{{\bf u}_b\cup w}}{
\Delta_{{\bf u}_b}\,.} 
\end{equation}

Let us note the range of indices in the above determinant formula. $(\alpha,a)$ takes values in the set 
\begin{equation}\label{Aset}
    \{(1,1),(1,2),(2,1),\dots,(L,2) \}
\end{equation}
whereas $(b,\beta)$ takes values in the set
\begin{equation}\label{Bset}
    \{(0,1),\dots,(0,L_0),(1,1),\dots,(1,L_1),\dots, (3,L_3) \}\,.
\end{equation}
Note that, in order to simplify this derivation, this notation is in contrast to the one used in the main text, where the rows of the determinant were labelled by $(a,\alpha)$ instead of $(\alpha,a)$. At the end we will convert back to the original ordering. 

In \cite{Gromov:2020fwh} a determinant relation was used to extract the SoV matrix elements for the measure, which in our notation corresponds to the case $L_0=L_3=0$ and $L_1=L_2=L$. For the general case we have the following updated determinant relation, valid for any two tensors $H_{a,\alpha,\beta}$ and $G_{a,\alpha,b}$, which reads 
\beq\label{detrelation2}
\det_{(\alpha,a),(b,\beta)}
H_{a,\alpha,\beta} G_{a,\alpha,b}=
\sum_{\sigma}(-1)^{|\sigma|}
\left(\prod_{b}
\det_{(\alpha,a)\in \sigma^{-1}(b),\beta_b}
H_{a,\alpha,\beta_b}\right)
\prod_{a,\alpha}
G_{a,\alpha,\sigma_{a,\alpha}}
\eeq
which is easy to derive. Here, $\sigma$ is a permutation of
\begin{equation}
    \{\underbrace{0,\dots,0}_{L_0},\dots,\underbrace{3,\dots,3}_{L_3}\}
\end{equation}
with $\sigma_{\alpha,a}$ denoting the number at position $a+(N-1)(\alpha-1)$ and 
\begin{equation}
    \sigma^{-1}(b) = \{(\alpha,a):\sigma_{a,\alpha}=b\}\,.
\end{equation}
We have $\beta_b\in \{1,\dots,L_b\}$ and finally $|\sigma|$ denotes the number of elementary permutations needed to bring the set $\bigcup_b \sigma^{-1}(b)$ to the canonical ordering \eqref{Aset}.

We now apply \eqref{detrelation2} to \eqref{integral3} by identifying
\begin{equation}
    H_{a,\alpha,\beta} = w_{\alpha,a}^{\beta-1}, \quad G_{a,\alpha,b} = f_b(w_{\alpha,a})Q_{1,1+a}\left(w_{\alpha,a}+\frac{i}{2}(3-2b) \right)\,.
\end{equation}
Notice that $\det_{(\alpha,a)\in \sigma^{-1}(b),\beta_b}
H_{a,\alpha,\beta_b}=(-1)^{\frac{L_b}{2}(L_b-1)}\Delta_b$
where $\Delta_b$ denotes the Vandermonde determinant built out of $w_{\alpha,a}$ for which $\sigma_{\alpha,a}=b$, that is 
\begin{equation}
    \Delta_b := \displaystyle\prod_{(\alpha,a)<(\alpha',a')}(w_{\alpha,a}-w_{\alpha',a'})
\end{equation}
where $<$ is to be understood in lexicographical ordering as explained above. The result then reads
\begin{equation}
  t(\{w_{\alpha,a}\}=  s'_{\bf L}\displaystyle \sum_{\sigma}(-1)^{\sigma|} \prod_b \Delta_b \prod_{\alpha,a} f_{\sigma_{a,\alpha}}(w_{\alpha,a})Q_{1,1+a}(w_{\alpha,a}+\tfrac{i}{2}+i s_{a,\alpha})
\end{equation}
where we have defined $s_{\alpha,a}=1-\sigma_{\alpha,a}$ and 
\begin{equation}
  s'_{\bf L}:=  \prod_b (-1)^{\frac{L_b}{2}(L_b-1)}
\end{equation}

Using the explicit form of $f_b(w)$, which is \eqref{fbdef}, it is easy to verify that
\begin{equation}
    \displaystyle\prod_b \Delta_b\prod_{\alpha,a}f_{\sigma_{a,\alpha}}(w_{\alpha,a}) = \prod_b \frac{\Delta_{{\bf u}_b\cup w_{\sigma^{-1}(b)}}}{
\Delta_{{\bf u}_b}}
\end{equation}
and hence we obtain 
\begin{equation}
    t(\{w_{\alpha,a}\})= s'_{\bf L}\displaystyle \sum_{\sigma}(-1)^{|\sigma|} \prod_b (-1)^{\frac{L_b}{2}(L_b-1)}\frac{\Delta_{{\bf u}_b\cup w_{\sigma^{-1}(b)}}}{
\Delta_{{\bf u}_b}} \prod_{\alpha,a} Q_{1,1+a}(w_{\alpha,a}+\tfrac{i}{2}+i s_{\alpha,a})\,.
\end{equation}
We now symmetrise\footnote{To avoid confusion, for any function $f(u,v)$ we define the symmetrisation of $f$ over $u$ and $v$ as being $\underset{u\leftrightarrow v}{\rm sym}\ f(u,v)= \frac{1}{2}\left(f(u,v)+f(v,u)\right)$, as used in Sections 5.2 and 5.3 of \cite{Gromov:2020fwh}.} over the integration variables $w_{\alpha,1}$ and $w_{\alpha,2}$. The only factor in \eqref{integral3} not invariant under this operation is $t(\{w_{\alpha,a}\})$, so symmetrising it  gives
\begin{equation}\label{symt}
        \underset{{w_{\alpha,1}\leftrightarrow w_{\alpha,2}}}{\rm sym}t(\{w_{\alpha,a}\}) =\frac{s'_{\bf L}}{2^L}\displaystyle \sum_{\sigma}(-1)^{|\sigma|} \prod_b \frac{\Delta_{{\bf u}_b\cup w_{\sigma^{-1}(b)}}}{
\Delta_{{\bf u}_b}} \prod_{\alpha} F_{\alpha}^{s_{\alpha,1}s_{\alpha,2}}\,,
\end{equation}
where 
\begin{equation}
    F_{\alpha}^{s_{\alpha,1}s_{\alpha,2}}=\det_{1\leq a,a'\leq 2}Q_{1,1+a}(w_{\alpha,a'}+\tfrac{i}{2}+i s_{\alpha,a'})\,.
\end{equation}
We will now derive this relation. We introduce the expression 
\begin{equation}
    h_{\sigma,\alpha}:=(-1)^{|\sigma|}\prod_b \frac{\Delta_{{\bf u}_b\cup w_{\sigma^{-1}(b)}}}{
\Delta_{{\bf u}_b}}Q_{1,2}(w_{\alpha,1}+\tfrac{i}{2}+i s_{\alpha,1})Q_{1,3}(w_{\alpha,2}+\tfrac{i}{2}+i s_{\alpha,2})\,.
\end{equation}
Consider the interchange of $w_{\alpha,1}$ and $w_{\alpha,2}$. This produces
\begin{equation}
h_{\sigma,\alpha}\rightarrow (-1)^{|\sigma|}\prod_b \frac{\bar{\Delta}_{{\bf u}_b\cup w_{\sigma^{-1}(b)}}}{
\Delta_{{\bf u}_b}}Q_{1,2}(w_{\alpha,2}+\tfrac{i}{2}+i s_{\alpha,1})Q_{1,3}(w_{\alpha,1}+\tfrac{i}{2}+i s_{\alpha,2})\,.
\end{equation}
where $\bar\Delta$ denotes that we have interchanged $w_{\alpha,1}$ and $w_{\alpha,2}$ inside the Vandermonde determinant. 

There are two possible types of $\sigma$. Either $\sigma_{\alpha,1}=\sigma_{\alpha,2}=c$ for some $c\in\{0,1,2,3\}$ or not. First suppose $\sigma_{\alpha,1}=\sigma_{\alpha,2}=c$. Then $s_{\alpha,1}=s_{\alpha,2}$ and, since $(\alpha,1)$ and $(\alpha,2)$ are adjacent to each other in the properly ordered set \eqref{Aset} we have 
\begin{equation}
    \bar{\Delta}_{{\bf u}_c\cup w_{\sigma^{-1}(c)}}=-\Delta_{{\bf u}_c\cup w_{\sigma^{-1}(c)}},\quad \bar{\Delta}_{{\bf u}_b\cup w_{\sigma^{-1}(b)}}=\Delta_{{\bf u}_b\cup w_{\sigma^{-1}(b)}},\ b\neq c\,.
\end{equation}
Hence, after exchanging $w_{\alpha,1}$ and $w_{\alpha,2}$ for such a $\sigma$ we obtain 
\begin{equation}
h_{\sigma,\alpha}\rightarrow -(-1)^{|\sigma|}\prod_b \frac{\Delta_{{\bf u}_b\cup w_{\sigma^{-1}(b)}}}{
\Delta_{{\bf u}_b}}Q_{1,2}(w_{\alpha,2}+\tfrac{i}{2}+i s_{\alpha,2})Q_{1,3}(w_{\alpha,1}+\tfrac{i}{2}+i s_{\alpha,1})\,.
\end{equation}
Hence for such a $\sigma$, after symmetrising over $w_{\alpha,1}$ and $w_{\alpha,2}$ we obtain
\begin{equation}
    \underset{{w_{\alpha,1}\leftrightarrow w_{\alpha,2}}}{\rm sym} h_{\sigma,\alpha} = \frac{1}{2}(-1)^{|\sigma|}\prod_b \frac{\Delta_{{\bf u}_b\cup w_{\sigma^{-1}(b)}}}{
\Delta_{{\bf u}_b}}F_{\alpha}^{s_{\alpha,1}s_{\alpha,2}}\,.
\end{equation}
We now consider the case of $\sigma$ such that $\sigma_{\alpha,1}=c_1$, $\sigma_{\alpha,2}=c_2$ with $c_1\neq c_2$. Note that there is another (unique) permutation $\tilde{\sigma}$ with $\tilde{\sigma}_{\alpha,1}=c_2$, $\tilde{\sigma}_{\alpha,2}=c_1$ and $\tilde{\sigma}_{\alpha',a'}=\sigma_{\alpha',a'}$ for all other pairs $(\alpha',a')$. Clearly since these two permutations are equivalent up to interchanging a single pair we have $(-1)^{|\sigma|}=-(-1)^{|\tilde{\sigma}|}$. Denote $\tilde{s}_{\alpha,a}=1-\tilde{\sigma}_{\alpha,a}$. Then it immediately follows that under exchanging $w_{\alpha,1}$ and $w_{\alpha,2}$ we have
\begin{equation}
    h_{\sigma,\alpha}\rightarrow -(-1)^{|\tilde{\sigma}|}\prod_b \frac{\Delta_{{\bf u}_b\cup w_{\tilde{\sigma}^{-1}(b)}}}{
\Delta_{{\bf u}_b}}Q_{1,2}(w_{\alpha,2}+\tfrac{i}{2}+i \tilde{s}_{\alpha,2})Q_{1,3}(w_{\alpha,2}+\tfrac{i}{2}+i \tilde{s}_{\alpha,1})\,.
\end{equation}
Hence, after symmetrisation we have
\begin{equation}
\underset{{w_{\alpha,1}\leftrightarrow w_{\alpha,2}}}{\rm sym} (h_{\sigma,\alpha}+h_{\tilde{\sigma},\alpha})= \frac{1}{2}\left((-1)^{|\sigma|}\prod_b \frac{\Delta_{{\bf u}_b\cup w_{\sigma^{-1}(b)}}}{
\Delta_{{\bf u}_b}} F_{\alpha}^{s_{\alpha,1}s_{\alpha,2}}+\sigma\leftrightarrow\tilde{\sigma}\right)\,.
\end{equation}
Of course, the conclusion is unchanged if $h_{\sigma,\alpha}$ is multiplied by any function independent of $(\alpha,a)$ and hence \eqref{symt} immediately follows by sequentially symmetrising over $(\alpha,1)$ and $(\alpha,2)$ for $\alpha=1,2,\dots,L$.

We now put \eqref{symt} under the integration \eqref{integral3} and compute the integral by residues, closing the contour in the upper-half plane. This produces a sum over poles at the locations $w_{\alpha,a}=\svx_{\alpha,a}=\theta_\alpha+i(\bs +n_{\alpha,a})$, with $n_{\alpha,a}$ ranging over all non-negative integers. If all $n_{\alpha,a}$ are distinct for a fixed $\alpha$ we can use the symmetry of the integrand to remove a factor of $2$ for each $\alpha$ and restrict the summation to $n_{\alpha,1}\geq n_{\alpha,2}$. If some $n_{\alpha,a}$ coincide for a fixed $\alpha$ then removing the $2^L$ factor will result in an overcounting which we must compensate for, by introducing the factor $M_\alpha$.

As a result, we obtain 
\begin{equation}
\begin{split}
     \sum_{n_{\alpha,1}\geq n_{\alpha,2}\geq 0}\displaystyle & \prod_{\alpha}\frac{1}{M_{\alpha}}\prod_{\alpha,a}\tilde{Q}_1^B(\svx_{\alpha,a})\frac{r_{\alpha,n_{\alpha,a}}}{r_{\alpha,0}}\\ 
    & s'_{\bf L}\sum_{\sigma}(-1)^{|\sigma|} \prod_b \frac{\Delta_{{\bf u}_b\cup \svx_{\sigma^{-1}(b)}}}{
\Delta_{{\bf u}_b}} \prod_{\alpha} \displaystyle \det_{1\leq a,a' \leq 2} Q_{1,a+1}(\svx_{\alpha,a'}+\tfrac{i}{2}+is_{\alpha,a})\,.
\end{split}
\end{equation}
We now compare with the general expression \eqref{genexpress}. We see that in order for a term
\begin{equation}\label{xymatelem}
        \Big[
L_0;{\bf u}_0
\Big|
L_1;{\bf u}_1
\Big|
L_2;{\bf u}_2
\Big|
L_3;{\bf u}_3
\Big]_{\svx\svy}
\end{equation}
in the summand of the r.h.s. to be non-zero it must be possible to write each $\svy_{\alpha,a}=\theta_\alpha+i(\bs+m_{\alpha,a}+1-a)$, for each fixed $\alpha$, as  
\begin{equation}
    \svy_{\alpha,a}=\svx_{\alpha,\rho^\alpha_a}+i s_{\alpha,\rho^{\alpha}_a,},\quad a=1,2
\end{equation}
where $\rho^{\alpha}$ is a permutation of $\{1,2\}$ and $\rho^\alpha_a:=\rho^\alpha(a)$ and hence we require
\begin{equation}\label{mnrelation}
    m_{\alpha,a} + 1-a =n_{\alpha,\rho^\alpha_a}+ s_{\alpha,\rho^{\alpha}_a}\,.
\end{equation}
Since each of the numbers  $m_{\alpha,a} + 1-a$ must be distinct, as otherwise the determinant built from $Q_{1,1+a}$ will vanish, there is a unique permutation $\rho^\alpha$ (if such a permutation exists) for which \eqref{mnrelation} holds. If such a permutation does not exist then the matrix element \eqref{xymatelem} vanishes. The permutation $\rho^\alpha$ amounts to sorting the set $\{n_{\alpha,1}+s_{\alpha,1},n_{\alpha,2}+s_{\alpha,2}\}$ and so we should keep track of the sign of this permutation. Hence, for a fixed permutation $\sigma$ we read off the following contribution to \eqref{xymatelem}
\begin{equation}\label{fixedsigma}
s'_{\bf L}(-1)^{\frac{L}{2}(L-1)(N-1)}\left.\frac{(-1)^{|\sigma|}}{\Delta_\theta^2} \prod_{\alpha}\frac{(-1)^{|\rho^\alpha|}}{M_{\alpha}}\prod_{\alpha,a}\frac{r_{\alpha,n_{\alpha,a}}}{r_{\alpha,0}}\prod_b(-1)^{\frac{L_b}{2}(L_b-1)} \frac{\Delta_{{\bf u}_b\cup \svx_{\sigma^{-1}(b)}}}{
\Delta_{{\bf u}_b}}\right|_{m_{\alpha,a}=n_{\alpha,\rho^\alpha_a}-\sigma_{\alpha,\rho^{\alpha}_a}+a} 
\end{equation}
where we have included the corresponding normalisation factor $\lN$. Finally, in order to determine \eqref{xymatelem} we note that for a given set of $\svx_{\alpha,a}$ and $\svy_{\alpha,a}$ there can be many different $\sigma$ for which the relation \eqref{mnrelation} holds and we must sum over all such $\sigma$ in \eqref{fixedsigma} in order to obtain \eqref{xymatelem}. When there is a degeneracy in $n_{\alpha,a}$ for fixed $\alpha$ there are multiple $\sigma$ that give the same result. Their number is exactly $M_\alpha$, so we can simplify the expression by only summing over $k$ inequivalent permutations of $n_{\alpha,a}$ within each $\alpha$. We denote such permutations ${\rm perm}_{\alpha}n$ and hence obtain 
\begin{equation}
     \langle \svy|\hat{O}|\svx\rangle =s_{\bf L}(-1)^{\frac{L}{2}(L-1)(N-1)} \left.\displaystyle \sum_{k} \frac{(-1)^{|\sigma|}}{\Delta_\theta^2}\prod_{\alpha,a}\frac{r_{\alpha,n_{\alpha,a}}}{r_{\alpha,0}} \prod_b (-1)^{\frac{L_b}{2}(L_b-1)}\frac{\Delta_{{\bf u}_b\cup \svx_{\sigma^{-1}(b)}}}{
\Delta_{{\bf u}_b}}\right|_{\sigma_{\alpha,a} = k_{\alpha,a}-m_{\alpha,a}+a}\,.
\end{equation}
Moving back to the original ordering of the rows of the determinant introduces a sign $(-1)^{\frac{L}{4}(N^2-3N+2)}$ which combines with $s'_{\bf L}$ to produce $s_{\bf L}$ given by 
\begin{equation}
    s_{\bf L}:=(-1)^{\frac{LN}{4}(L-1)(N-1)+\sum_{n=0}^N L_n}\,.
\end{equation}
Finally, the above argument is rigourous in the finite-dimensional setting. To pass to the infinite-dimensional case we notice that the matrix elements are block diagonal with each block having finite size. The spin $\bs$ enters each block as a universal polynomial pre-factor. Then, each block is fixed by analysing a finite-number of finite-dimensional representations and the matrix elements can be analytically continued to values of $\bs$ not being negative half-integers. So the matrix elements we found are valid in the infinite-dimensional case as well for generic $\bs$. Note that since the SoV basis vectors are polynomial in the spin $\bs$ our formula for the SoV matrix elements of the principal operators are valid for all values of $\bs$.

This completes the derivation. The $\sl(N)$ case is identical up to extending the range of indices $\{1,2\}$ to $\{1,\dots,N-1\}$ but can be carried out in exactly the same way as was demonstrated for the measure in \cite{Gromov:2020fwh}. Note that for a function $f(w_{\alpha,1},\dots,w_{\alpha,N-1})$ we define the symmetrisation over $w_{\alpha,1},\dots,w_{\alpha,N-1}$ as 
\begin{equation}
    \frac{1}{(N-1)!}\sum_{p\in\mathfrak{\sigma}_{N-1}}f(w_{\alpha,p(1)},\dots, w_{\alpha,p(N-1)})
\end{equation}
where $\mathfrak{\sigma}_{N-1}$ denotes the permutation group on $N-1$ letters. 

\section{SoV basis}\label{deriveB}

In this section we will demonstrate that knowledge of the structure of the SoV basis and the FSoV approach allows one to derive the form of Sklyanin's ${\bf B}$ operator.  

We start by defining the SoV ground states $|0\rangle$ and $\langle 0|$ which correspond to the constant polynomial $1$. These states satisfy the following properties
\begin{equation}\label{SoVvacprop}
    T_{j1}(\theta_\alpha+i\bs)|0\rangle = 0=T_{1j}(\theta_\alpha+i\bs)|0\rangle = 0,\quad j=1,\dots,N,\quad \alpha=1,\dots,L\,.
\end{equation}
We can now follow the logic of \cite{Maillet:2018bim} and build vectors by action of transfer matrices on $\langle 0|$ and $|0\rangle$. The key idea of \cite{Maillet:2018bim} is that if such vectors form a basis then it is automatically an SoV basis since the transfer matrix wave functions will factorise. We will choose the following set of transfer matrices
\begin{equation}
   \bbT^*_{\mu}(u):=\det_{1\leq j,k\leq \mu_1}\bbT_{\mu'_j-j+k,1}\left(u-\frac{i}{2}\left(\mu'_1-\mu_1-\mu'_j+j+k-1\right)\right)
\end{equation}
where $\T_{a,1}(u)$ are the transfer matrices in anti-symmetric representations \eqref{highertransfer} and $\mu$ denotes an integer partition (Young diagram)
\begin{equation}
    \mu=(\mu_1,\dots,\mu_{N-1},0)
\end{equation}
and $\mu'_j$ denotes the height of the $j$-th column of the Young diagram. The states $|\svy\rangle$ are then constructed as 
\begin{equation}\label{cvecs}
 |\svy\rangle\, \propto\,  \prod_{\alpha=1}^L \bbT^*_{\mu^\alpha}\left(\theta_\alpha+i\bs+\frac{i}{2}\left(\mu_1^\alpha-\mu^{\alpha,\prime}_1\right)\right)|0\rangle
\end{equation}
and we label the constructed states by the $L$ Young diagrams $\mu^\alpha$, $\alpha=1,\dots,L$. 

We also construct a set of left vectors 
\begin{equation}\label{bvecs}
   \langle\svx|\, \propto \, \bra{0}\prod_{\alpha=1}^L\prod_{j=1}^{N-1}\bbT_{N-1,s_j^\alpha}\left(\theta_\alpha+i \bs-\frac{i}{2}(N-s^\alpha_j-1)\right)
\end{equation}
where $(N-1,s)$ denotes the Young diagram of height $N-1$ and width $s$, that is $\mu=(\underbrace{s,\dots,s}_{N-1},0)$ and the corresponding transfer matrix is defined by the Cherednik-Bazhanov-Reshetikhin (CBR) \cite{Cherednik,Bazhanov:1989yk} formula 
\begin{equation}
    {\mathbb T}_{\mu}(u)=\det_{1\leq j,k\leq \mu_1}{\mathbb  T}_{\mu'_j-j+k,1}\left(u-\frac{i}{2}\left(\mu'_1-\mu_1-\mu'_j+j+k-1\right)\right)\;.
\end{equation}

We now note two key properties of the constructed set of vectors. First, they are linearly independent. This was proven in \cite{Ryan:2020rfk} and the argument relies on the fact that the twist matrix \eqref{companion} can be deformed slightly with $N-1$ parameters $w_1,\dots,w_{N-1}$. Then, in the limit where all $w_i$ are sent sequentially to infinity the constructed set of vectors reduce to eigenvectors of the so-called Gelfand-Tsetlin algebra \cite{Ryan:2020rfk}, a key object in representation theory. Furthermore, the Gelfand-Tsetlin algebra has non-degenerate spectrum and it was shown in \cite{Ryan:2020rfk} that a basis of eigenvectors are given by \eqref{bvecs} and \eqref{cvecs} in the above-described limit. Hence, the vectors \eqref{bvecs} and \eqref{cvecs} form a basis, and the transfer matrix wave functions are guaranteed to factorise.

The next key property is that the constructed set of vectors are independent of the twist eigenvalues. This follows from the fact that 
all transfer matrices in our chosen reference frame have the structure 
\begin{equation}
    \T^*_\mu(u) = \T_\mu^{*,0}(u) + \sum_{r=0}^N \dots\times\chi_r T_{r1}(u)
\end{equation}
and 
\begin{equation}
    \T_{N-1,s}(u) = \T_{N-1,s}^{0}(u) + \sum_{r=0}^N \chi_r T_{1,r}(u)\times\dots
\end{equation}
where $\T_\mu^{0}(u)$ denotes a part which is independent of the twist eigenvalues. The property \eqref{SoVvacprop} then ensures that the twist-dependent part of the transfer matrices never contributes, see \cite{Ryan:2018fyo,Ryan:2020rfk,Gromov:2020fwh}. 

We now exploit known relations for transfer matrices in terms of Baxter Q-functions. The transfer matrix eigenvalues admit the form
\begin{equation}
    \bra{\Psi}\bbT^*_{\mu^\alpha}\left(\theta_\alpha+i\bs+\frac{i}{2}\left(\mu_1^\alpha-\mu^{\alpha,\prime}_1\right)\right) \ \propto\  \frac{\displaystyle\det_{1\leq a,a'\leq N-1} Q^{a+1} \left(\svy_{\alpha,a'}+\frac{i}{2}\left(N-2\right)\right)}{\displaystyle \det_{1\leq a,a'\leq N-1} Q^{a+1}\left(\theta_\alpha+i\bs + \frac{i}{2}\left(N-2k\right)\right)}\bra{\Psi}
\end{equation}
with $\svy_{\alpha,a}= \theta_\alpha+i(\bs+\mu^\alpha_a+1-a)$ and 
\begin{equation}
    \prod_{\alpha=1}^L\prod_{a=1}^{N-1}\bbT_{N-1,s_a^\alpha}\left(\theta_\alpha+i \bs-\frac{i}{2}(N-s^\alpha_j-1)\right)|\Psi\rangle \, \propto\, \prod_{\alpha=1}^L \prod_{a=1}^{N-1} \frac{Q_1(\theta_\alpha+i\bs +i s_a^\alpha )}{Q_1(\svx_{\alpha,a})}|\Psi\rangle
\end{equation}
where $\svx_{\alpha,a}=\theta_\alpha+i(\bs+s^\alpha_a)$. 

We can now write down the wave functions. By normalising $\langle\Psi|$ and $|\Psi\rangle$ appropriately we have
\begin{equation}
     \langle\Psi|\svy\rangle = \prod_{\alpha=1}^L \displaystyle\det_{1\leq a,a'\leq N-1} Q^{j+1} \left(\svy_{\alpha,a}+\frac{i}{2}\left(N-2\right)\right)\,.
\end{equation}
Similarly, we have
\begin{equation}
    \langle\svx|\Psi\rangle = \prod_{\alpha=1}^L \prod_{a=1}^{N-1} Q_1(\svx_{\alpha,a})
\end{equation}
Since the proposed sets of vectors form a basis we can write the scalar product between two transfer matrix eigenstates as 
\begin{equation}
    \langle\Psi_A|\Psi_B\rangle = \sum_{\svx,\svy} 
\Psi_B(\svx)\lM_{\svy,\svx} \Psi_A(\svy)\,.
\end{equation}
We now turn to the FSoV construction which allows us to extract the measure \eqref{measure} in the two SoV bases. This is just a special case of the formula \eqref{psixy}. Since the SoV bases are independent of twist, the character projection trick is valid and all of the techniques developed earlier in the paper in sections \ref{sec:sl3disc} and \ref{sec5} can be carried out. In particular, we can compute correlation functions of multi-insertions of principal operators. Following the logic of section \ref{secBC} we see that there is a distinguished operator diagonalised in the basis $|\svx\rangle$ which then must also be diagonalised in the basis $\langle \svx|$ defined in \eqref{bvecs}. Hence, we have obtained Sklyanin's ${\bf B}$ operator, and the basis diagonalising it, starting solely from the FSoV approach and the knowledge of the SoV basis. 

\section{Existence of basis of Bethe algebra eigenstates}\label{basisapp}

In this appendix we will prove the existence of the decomposition used in \eqref{complrel} which states that one can write a resolution of the identity as 
\begin{equation}\label{decompeigen}
    1 = \displaystyle\sum_{A}|\Psi_A\rangle\langle \bar{\Psi}_A|
\end{equation}
where each $|\Psi_A\rangle$ is a joint eigenvector of the transfer matrices $t_1(u)\, , t_2(u)$, $\langle \bar{\Psi}_A|$ is defined by the property 
\begin{equation}
    \langle\bar{\Psi}_A|\Psi_B\rangle=\delta_{AB}
\end{equation}
and the index $A$ in the sum labels all transfer matrix eigenstates. 
\paragraph{Outline of proof}
The proof is very similar in spirit to our proof in the paper \cite{Gromov:2020fwh} showing that eigenstates of the SoV $\bB$ operator form a basis of the representation space. An outline of the steps we will take are as follows. We will begin by decomposing the representation space $\mathcal{H}$ into a direct sum of finite-dimensional spaces $\mathcal{H}_k$, $k\in \mathbb{Z}_{\geq 0}$ from which the identity operator inherits the decomposition 
\begin{equation}
1 = \displaystyle \sum_{k\geq 0} 1_k    
\end{equation}
where $1_k$ denotes the restriction of the identity operator to $\mathcal{H}_k$. We will then prove that each $\mathcal{H}_k$ admits a basis of transfer matrix eigenstates and as a result we can write 
\begin{equation}
    1_k = \displaystyle\sum_{A_k}|\Psi_{A_k}\rangle\langle \bar{\Psi}_{A_k}|
\end{equation}
where the sum is over a finite number of certain transfer matrix eigenstates enumerated by $A_k$ to be specified later. Hence, the decomposition \eqref{decompeigen} holds with 
\begin{equation}
    1 = \displaystyle\sum_{k\geq 0} \displaystyle\sum_{A_k}|\Psi_{A_k}\rangle\langle \bar{\Psi}_{A_k}|\,.
\end{equation}

\paragraph{Proof}

The representation space of the $\sl(3)$ spin chain is the space of polynomials in $x_\alpha,\, y_\alpha$, $\alpha=1,\dots,L$. By definition, a vector in this space is a finite linear combination of monomials 
\begin{equation}
\displaystyle\prod_{\alpha=1}^L x_\alpha^{n_\alpha} y_\alpha^{m_\alpha},\quad  n_\alpha,m_\alpha\in \mathbb{Z}_{\geq 0}\,.
\end{equation}
In order to perform $\mathsf{SL}(3)$ (group)-valued linear transformations the representation space must be extended from polynomials to analytic functions which can be represented as power series in the above variables convergent in some neighbourhood of the origin, see \cite{Gromov:2020fwh} for a discussion and examples. Hence, in order to show that a collection of vectors form a basis of the representation space we must show that any such analytic function can be written as an infinite series in those vectors with finite coefficients.

Note that there are legitimate questions about convergence of such infinite series and the existence of the corresponding finite coefficients. Indeed even if one manages to construct a basis for the original space of polynomials it does not necessarily extend to a basis of the space of analytic functions.

Let us introduce the following operator $\mathcal{E}=-\mathcal{E}_{11}-L\,\bs\times 1$ of the global Cartan subalgebra of $U(\gl(3))$, see \eqref{globalops22} for the definition of $\mathcal{E}_{11}$. A direct calculation yields
\begin{equation}
    \mathcal{E} \displaystyle\prod_{\alpha=1}^L x_\alpha^{n_\alpha} y_\alpha^{m_\alpha}= \left(\displaystyle\sum_{\alpha=1}^L n_\alpha+m_\alpha\right)\displaystyle\prod_{\alpha=1}^L x_\alpha^{n_\alpha} y_\alpha^{m_\alpha}\,.
\end{equation}
Hence, we see that the spectrum of this operator is bounded from below and furthermore each eigenvalue is non-negative. Hence the representation space $\mathcal{H}$ decomposes into a direct sum of eigenspaces $\mathcal{H}_k$ corresponding to the eigenvalue $k$
\begin{equation}\label{Hilbdecomp}
    \mathcal{H}= \underset{k\geq 0}{\bigoplus}\, \mathcal{H}_{k},\quad \mathcal{H}_k:=\{v\in \mathcal{H}\, :\, \mathcal{E}v = k\, v \}\,.
\end{equation}
Clearly, each $\mathcal{H}_k$ is finite-dimensional as there are only finitely many ways to write a given non-negative integer $k$ as a sum of non-negative integers $n_\alpha$ and $m_\alpha$. Clearly the decomposition \eqref{Hilbdecomp} is valid for the space of polynomials. However, since the space of regular at the origin analytic functions admit a Taylor expansion into the variables $x_\alpha,y_\alpha$ the decomposition \eqref{Hilbdecomp} is also valid for this enlarged space. To complete the proof it remains to show that each $\lH_k$ admits a basis of transfer matrix eigenstates. To this end it suffices to show that they have a distinct set of eigenvalues in each subspace the number of which matches the dimension of that space. Since transfer matrix eigenvalues are algebraic functions of the twist parameters it suffices to prove there are sufficiently many for some special value of the twist to prove that it is true generically.

We will proceed as follows. Consider the transfer matrices with diagonal twist $g={\rm diag}(\lambda_1,\lambda_2,\lambda_3)$. As a result the transfer matrices commute with $\mathcal{E}$ and preserve the subspaces $\lH_k$.  We consider the singular twist limit $\lambda_1\gg \lambda_2\gg\lambda_3$. In this limit $t_1(u)$ and $t_2(u)$ reduce to the generators $\mathsf{GT}_1(u)$ and $\mathsf{GT}_2(u)$ of the Gelfand-Tsetlin subalgebra of the Yangian whose properties are well-understood \cite{molev2007yangians}. Let $\sigma\in\mathbb{C}$ be a generic parameter and consider the combination 
\begin{equation}
    t(u) = \frac{1}{\lambda_1}t_1(u) + \frac{\sigma}{\lambda_1\lambda_2}t_2(u)\,.
\end{equation}
In the singular twist limit $t(u)\rightarrow t^{\mathsf{GT}}(u):= \mathsf{GT}_1(u) + \sigma\, \mathsf{GT}_2(u)$. We will prove that this operator has distinct eigenstates in each Cartan subspace, and hence so does $t(u)$ and hence so do the family $t_1(u)$ and $t_2(u)$. 

The eigenvectors of $t^{\mathsf{GT}}(u)$ are well-understood for finite-dimensional representations where $\bs\in\{0,-\frac{1}{2},-1,\dots\}$\,. In particular, they are all polynomial functions of the spin $\bs$ with each eigenvalue also being a polynomial in $\bs$. What is not obvious is that each eigenvector remains an eigenvectors when analytically continued to non half-integer values of $\bs$. Let us consider the expression  
\begin{equation}\label{anaspins}
    t^{\mathsf{GT}}(u)|\Psi_\bs\rangle-\tau^{\mathsf{GT}}(u)|\Psi_\bs\rangle
\end{equation}
where $|\Psi_\bs\rangle$ is an eigenvectors of $t^{\mathsf{GT}}(u)$ for $\bs\in\{0,-\frac{1}{2},-1,\dots\}$. For such values of $\bs$ this expression equates to zero. However, for generic $\bs$ the operator $t^{\mathsf{GT}}(u)$ is a differential operator with coefficients which are polynomial in $\bs$ and hence the action of $t^{\mathsf{GT}}(u)$ on $|\Psi_\bs\rangle$ results in a vector which is polynomial in $\bs$ and hence \eqref{anaspins} vanishes for all values of $\bs$. Hence, $|\Psi_\bs\rangle$ is an eigenvector for all $\bs$ and is non-zero for generic $\bs$ and $t^{\mathsf{GT}}(u)$ has distinct eigenvalue for each such eigenvector. Since each $\lH_k$ can be obtained by considering enough finite-dimensional representations with sufficiently large $-\bs$ we can promote a basis of eigenvectors of $\lH_k$ for $\bs\in \{0,-\frac{1}{2},\dots\}$ to a basis of eigenvectors for generic $\bs$. This completes the proof.

\end{document}